\documentclass{aa} 
\usepackage[varg]{txfonts}
\usepackage{color}
\usepackage{supertabular,lscape}
\usepackage{comment}
\usepackage{subcaption}
\newcommand{\kms} {km\,s$^{-1}$}

\newcommand{\Msun} {$\mbox{M}_{\sun}$}
\newcommand{\Mstar} {$\mbox{M}^{\star}$}

\newcommand{\oii}{[O\,{\sc ii}]$\,\lambda\lambda$3726,3729}
\newcommand{\ciii}{C\,{\sc iii}]$\,\lambda\lambda$1907,1909}
\newcommand{\lya}{Ly$\alpha$}

\begin{document}

\title{New criteria for the selection of galaxy close pairs from cosmological simulations: evolution of the major and minor merger fraction in MUSE deep fields\thanks{Based on observations made with ESO telescopes at the Paranal Observatory under programmes 094.A-0289, 094.A-0115, 094.A-0247, 095.A-0118, 095.A-0010, and 096.A-0045}
}

\author{E.Ventou\inst{1}
\and T. Contini\inst{1}
\and N. Bouché\inst{1,2}
\and B. Epinat\inst{1,3}
\and J. Brinchmann\inst{4,5}
\and H. Inami\inst{6,2}
\and J. Richard\inst{2}
\and I. Schroetter\inst{1,7}
\and G. Soucail\inst{1}
\and M. Steinmetz\inst{8}
\and P.M. Weilbacher\inst{8}
}
\institute{Institut de Recherche en Astrophysique et Planétologie (IRAP), Université de Toulouse, CNRS, UPS, F-31400 Toulouse, France
\and Univ Lyon, Univ Lyon1, Ens de Lyon, CNRS, Centre de Recherche Astrophysique de Lyon UMR5574, F-69230, Saint-Genis-Laval, France
\and Aix Marseille Univ, CNRS, CNES, LAM, Marseille, France
\and Instituto de Astrof{\'\i}sica e Ci{\^e}ncias do Espaço, Universidade do Porto, CAUP, Rua das Estrelas, PT4150-762 Porto, Portugal
\and Leiden Observatory, Leiden University, P.O. Box 9513, 2300 RA, Leiden, The Netherlands
\and Hiroshima Astrophysical Science Center, Hiroshima University, 1-3-1 Kagamiyama, Higashi-Hiroshima, Hiroshima 739-8526, Japan
\and GEPI, Observatoire de Paris, PSL Université, CNRS, 5 Place Jules Janssen, 92190 Meudon, France
\and Leibniz-Institut f\"ur Astrophysik Potsdam (AIP), An der Sternwarte 16, 14482 Potsdam, Germany
}

\date{Received date}

\abstract
{  
It is still a challenge to assess the merger fraction of galaxies at different cosmic epochs in order to probe the evolution of their mass assembly.
%Assessing the merger fraction of galaxies at different cosmic epochs is still a challenge in order to probe the evolution of mass assembly of galaxies.  
%Recent close pair count studies predict that the galaxy major merger fraction turn over or flatten for $z \geq 3$ which is in disagreement with other published results. 
Using the \textsc{Illustris} cosmological simulation project, we investigate the relation between the separation of galaxies in a pair, both in velocity and projected spatial separation space, 
%velocity-distance relative separation of galaxies in a pair 
and the probability that these interacting galaxies will merge in the future. From this analysis, we propose a new set of criteria to select close pairs of galaxies along with a new corrective term to be applied to the computation of the galaxy merger fraction. We then probe the evolution of the major and minor merger fraction using the latest MUSE deep observations over the Hubble Ultra Deep Field, Hubble Deep Field South, COSMOS-Gr30 and Abell 2744 regions. From a parent sample of 2483 galaxies with spectroscopic redshifts, we identify 366 close pairs spread over a large range of redshifts ($0.2<z<6$) and stellar masses ($10^7-10^{11}$\Msun).
%Among a spectroscopic parent sample of 2483 galaxies, 366 close pairs were identified spread over a large range of redshifts ($0.2<z<6$) and stellar masses ($10^7-10^{11}$\Msun). 
Using the stellar mass ratio between the secondary and primary galaxy as a proxy to split the sample into major, minor and very minor mergers, we found a total of 183 major, 142 minor and 47 very minor close pairs corresponding to a mass ratio range of 1:1-1:6, 1:6-1:100 and lower than 1:100, respectively. 
%Our estimates show the impact of dense environments such as galaxy cluster or group on the close pair counts. We find an enhancement of the major merger fraction in A2744 and COSMOS-Gr30 compared to lower density fields in the UDF-Mosaic or HDFS at similar redshifts.
Due to completeness issues, we do not consider the very minor pairs in the analysis. Overall, the major merger fraction increases up to $z\approx 2-3$ reaching 25\% for pairs with the most massive galaxy 
%for primary galaxies 
with a stellar mass \Mstar\ $\geq 10^{9.5}$\Msun. Beyond this redshift, the fraction decreases down to $\sim 5$\% at $z\approx 6$. The major merger fraction for lower mass primary galaxies \Mstar\ $\leq 10^{9.5}$\Msun, seems to follow a more constant evolutionary trend with redshift. Thanks to the addition of new MUSE fields and new selection criteria, the increased statistics of the pair samples allow to narrow significantly the error bars compared to our previous analysis (Ventou et al.\,2017). The evolution of the minor merger fraction is roughly constant with cosmic time, with a fraction of 20\% at $z<3$ and a slow decrease between $3\leq z \leq6$ to $8-13$\%.}

\titlerunning{Merger fraction in MUSE deep fields}
\authorrunning{E.\,Ventou et al.}

 \maketitle

\section{Introduction}
\label{sec:intro}

 \begin{figure*}[t]
	\includegraphics[width=\textwidth]{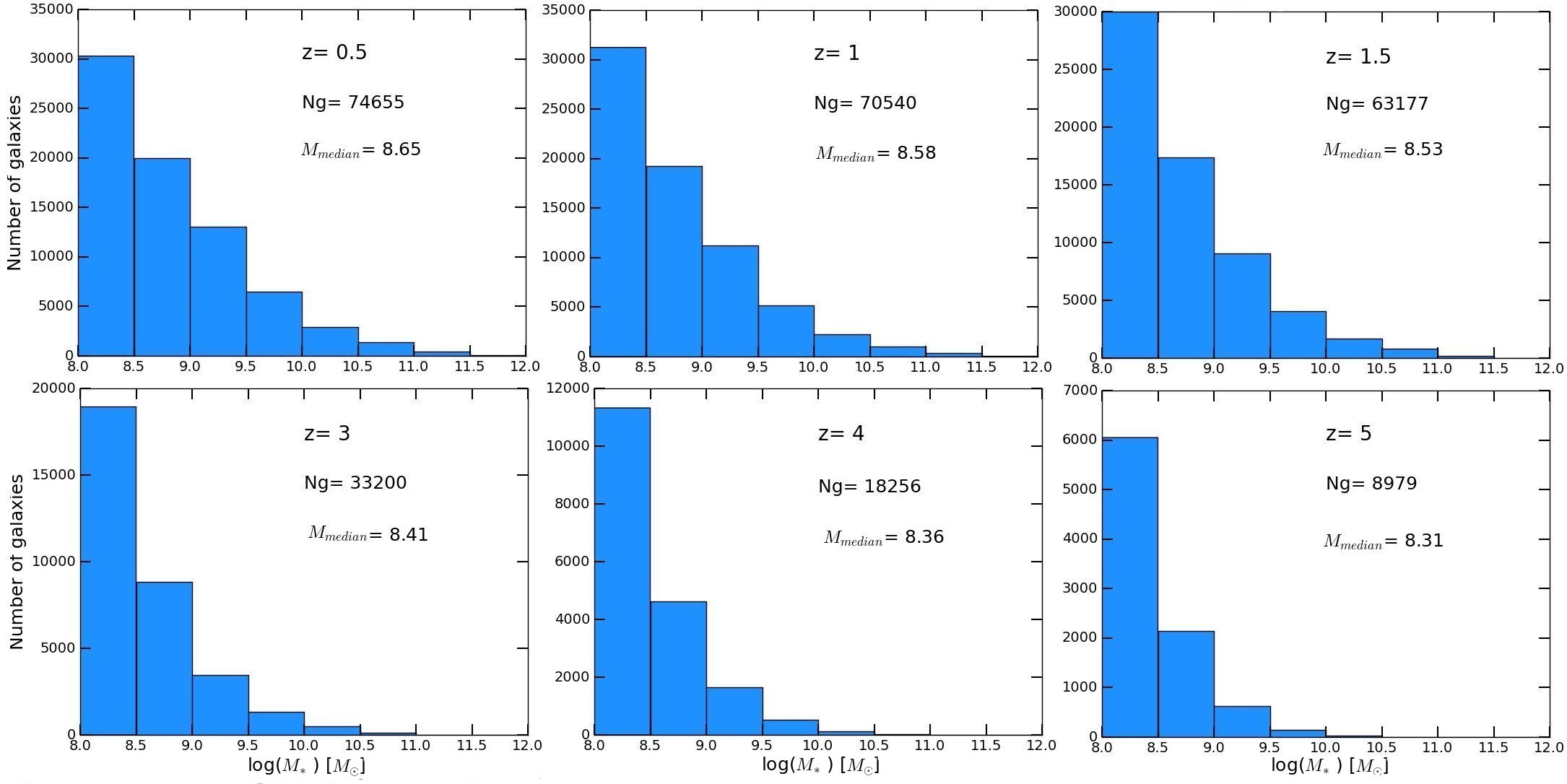}
   \caption{Stellar mass distribution of galaxies in the \textsc{Illustris} simulation for six snapshots corresponding to redshifts $z=$ 0.5, 1, 1.5, 3, 4 and 5. The total number of galaxies and the median stellar mass are indicated in each panel.}
	\label{fig:massdistrib_illustris}
\end{figure*}

Understanding the processes behind the mass assembly of galaxies in dark matter halos remains one of the most outstanding issues of modern astrophysics. Thanks to the development of more and more sophisticated cosmological models and simulations as well as new data coming from deep and wide photometric and spectroscopic surveys, much progress has been made both on the theoretical and observational side of galaxy evolution. Several mechanisms, such as cold gas accretion and galaxy mergers, contribute to the build-up of galaxies along cosmic time (e.g.\,Angl\'es-Alc\'azar et al. 2017).
In the first scenario, fresh gas is supplied to the galaxy from cold filaments following the cosmic web of large-scale structure. While direct observational evidence of this phenomenon has been difficult to obtain, indirect arguments, such as evidences from internal kinematic, expected absorption features along background quasar sight-lines or chemical evolution models with the well known G-dwarf problem, has been accumulating over the past decade (Chiappini 2001; Caimmi 2008; Sancisi et al. 2008; Bournaud et al. 2011; Stewart et al. 2011; Bouch\'e et al. 2016; Zabl et al. 2019).
In comparison many examples of colliding and merging galaxies have been observed and studied in the local universe. Galaxy mergers are known to not only enhance star formation and fuel starbursts (Joseph \& Wright 1985; Di Matteo et al. 2007; Kaviraj 2014), but also to strongly affect galaxy morphologies and dynamics (Bell et al. 2008; Perret et al. 2014; Borlaff et al. 2014; Lagos et al. 2018).  

The relative contribution and efficiency of these processes to the mass growth of galaxies is still unclear. Cosmological simulations suggest that a large fraction of cold gas can be accreted by galaxies and smooth gas accretion may dominate galaxy assembly, at least for low-mass galaxies hosted by halos below the virial shock mass threshold (Murali et al. 2002; Keres et al. 2005; Williams et al. 2011; L'Huillier et al. 2012; van de Voort et al. 2012; Conselice et al. 2013). The efficiency of smooth accretion onto halos seems to be more efficient at high redshift and increases slowly with the halo mass (e.g. Keres et al. 2005; Ocvirk et al. 2008; van de Voort et al. 2011). The same redshift dependence is seen in simulations for the cold gas accretion on to galaxies, but the dependence on halo mass is not constant and peaks around ($M_h \sim 10^{11-12}$ \Msun).

However, the relative importance of both phenomena (cold accretion and mergers) remains uncertain, since the total amount of mass accretion onto galaxies by merging is still poorly constrained, especially in the early epoch of galaxy evolution due to the difficulties to observe these events at high redshift.

Several methods have been used to investigate merging activity across cosmic time, for instance by identifying mergers through perturbations in galaxy morphologies 
%like CAS (concentration, C; asymmetry, A; clumpiness, S), Gini or $M_{20}$ parameter techniques 
(Le F\`evre et al. 2000; Conselice et al. 2003; Conselice 2006; Kampczyk et al. 2007; Conselice et al. 2008;   Heiderman et al. 2009; Bluck et al. 2012; Casteels et al. 2014).
%However these approaches are limited by the poor spatial resolution of high redshift objects, even with HST images, and to the fact that morphological disturbances are not always related to merger events, as suggested by galaxy kinematics (e.g.\, Förster Schreiber et al. 2009, 2011). 
However, such approaches are limited by the poor spatial resolution of broad band observations, by the insufficient depth of better resolution data, necessary to identify low surface brightness features such as tidal tails, and by the lack of near-infrared observations for large samples of high redshift galaxies. In addition, morphological disturbances are not necessarily related to merger events (e.g. Cibinel et al. 2015; Förster-Schreiber et al. 2009, 2011). Kinematics studies and numerical simulations have shown that a rotating disc can quickly ($< 1$ Gyr) rebuild after a merger at high redshift and that the signatures of a merger event are usually visible during a period of a few 100 Myr only (e.g. Perret 2014; Cibinel et al. 2015; Hung et al. 2016; Simons et al. 2019).
It has been shown that identifying mergers based on kinematics only is not straightforward and depends on the method used (e.g. Neichel et al. 2008; Shapiro et al. 2008; Epinat et al. 2012). The most secure way to study merging activity is thus to identify close pairs based on their kinematics.

At high redshift ($z \geq 2$),  studies  have thus  focused on the close pair counts method to probe merger abundance. These close pairs are  gravitationally bound systems of two galaxies and are expected to merge within an estimated timescale of about 1 Gyr (Kitzbichler \& White 2008; Jian et al. 2012; Moreno et al 2013) for nearly equal-mass galaxies (major merger with a mass ratio between the two galaxies greater than 1:4).

Several photometric and spectroscopic surveys have found that the major merger fraction and rate increase with redshift up to $z \sim 1$  (Lin et al. 2008; Bundy et al. 2009; de Ravel et al. 2009; Lopez-Sanjuan et al. 2012). Only a few estimates of this fraction and rate have been attempted for $z \geq 1.5$ and the conclusions reached on their evolution across cosmic time depend strongly on the adopted selection method.
Photometric and flux-ratio-selected major pairs studies reveal that the major merger rate increases steadily up to $z\sim 2-3$ (Bluck et al. 2009; Man et al. 2012, 2016; Lopez-Sanjuan et al. 2015; Mundy et al. 2017; Duncan et al. 2019), but see Mantha et al.\,(2018) for a contradictory result, whereas spectroscopic and mass-ratio-selected pairs from recent surveys found that beyond $z \geq 2$, the incidence of major mergers remains constant or decreases at early times (Lopez-Sanjuan et al. 2013; Tasca et al. 2014; Ventou et al. 2017).
This discrepancy could be explained by the contamination of photometric samples by a large number of minor mergers, with a mass ratio lower than 1:4, (Lotz et al. 2011; Mantha et al. 2018).
The large scatter between measurements using the same selection method can also be attributed to the wide range of companion selection criteria used in previous surveys. While it would be more accurate to identify close pairs of galaxies based on their true (i.e.\, in real space) physical separation, it is not applicable directly to the observed datasets. Thereby various criteria have been considered. 

The first analyses of galaxy pairs formulated a criterion mostly relying on apparent angular separation and angular diameter of the galaxies (Turner 1976a; Peterson 1979). In more recent sudies (Patton et al. 2000; de Ravel et al. 2009; Tasca et al. 2014; Man et al. 2016; Ventou et al. 2017) close pairs of galaxies are frequently defined as two galaxies within a limited projected angular separation and line-of-sight relative velocity. For spectroscopic surveys a relative velocity difference of $ \Delta_{V} \leqslant 300-500$ \kms\ is often applied, which offers a good compromise between contamination by chance pairing, i.e.\, pairs which will satisfy the selection criteria but are not gravitationally bound, and pair statistics (Patton et al. 2000;  Lin et al. 2008). The projected separation criterion however varies a lot in the literature, $0-10 \leqslant r_{p} \leqslant 25-50$ h$^{-1}$kpc, which makes direct comparisons difficult (Patton et al. 2000; de Ravel et al. 2009; Lopez-Sanjuan et al. 2013; Tasca et al. 2014; Ventou et al. 2017; Mantha et al. 2018).
Furthermore, recent studies based on the Sloan Digital Sky Survey have shown that effect of galaxy interactions can be detected in galaxy pairs with separation greater than 50 h$^{-1}$kpc.  Star formation rate (SFR) enhancements, for example, are present out to projected separations of 150 h$^{-1}$kpc (Scudder et al. 2012; Patton et al. 2013). This shows the need to investigate and better refine companion selection criteria for close pair count study.

While major mergers are  relatively easy to identify, minor mergers are more frequent in the nearby universe and may also be an important driver of galaxy evolution (Naab et al. 2009; McLure et al. 2013; Kaviraj 2014). However the cosmic evolution of the minor merger fraction and rate of galaxies is almost unconstrained, with very few attempts so far (e.g.\,Lopez-Sanjuan et al. 2011, 2012).

In the present paper, we aim to provide new selection criteria for close pair count analysis. We make use of the \textsc{Illustris} cosmological simulation project to investigate the relation between close pair selection criteria, i.e.\, separation distance and relative  velocity, and whether the two galaxies will finally merge by $z=0$. Following the analysis of Ventou et al. (2017), we apply these  new criteria to  MUSE (Multi-Unit Spectroscopic Explorer) deep observations performed over four regions: the Hubble Deep Field South, the Hubble Ultra Deep Field, the galaxy cluster Abell 2744, and the COSMOS-Gr30 galaxy group, in order to better constrain the cosmic evolution of the merger fraction. Thanks to its large field-of-view, MUSE allows  to explore the close environment of galaxies and thus to probe the evolution of the major and minor merger fraction over a wide range of stellar masses and redshift domain.

This paper is organized as follows: In section \ref{Illustris_data}, we introduce the \textsc{Illustris} simulation and we detail in section \ref{criteria} the analysis performed on the companion selection criteria and its results. The MUSE data sets used to detect galaxy close pairs as well as the final close pairs sample, are described in section \ref{sample}. Finally we give an estimate of the major and minor merger fraction evolution up to $z\sim 6$ in section \ref{merger_fraction}.
A summary and conclusion are given in section\,\ref{conclusion}.

Throughout this work, we use a standard $\Lambda$CDM cosmology with  $H_0=100 h$ kms$^{-1}$ Mpc$^{-1}$, $h=0.7$, $\Omega_{m}= 0.3$, $ \Omega_{\Lambda}= 0.7$. Magnitudes are in given in the AB system. 

\section{\textsc{Illustris} simulation}
\label{Illustris_data}

The \textsc{Illustris} cosmological simulation project (Vogelsberger et al. 2014; Genel et al. 2014; Nelson et al. 2015), is a series of N-body/hydrodynamical simulations reproducing the formation and evolution of galaxies across cosmic time over a large volume of $106.5\ \rm{Mpc}^3$.
The simulations uses the moving-mesh code AREPO (Springel 2010) and includes many ingredients for galaxy evolution such as primordial and metal-line cooling with self-shielding corrections, stellar evolution, stellar feedback, galactic-scale outflows with an energy-driven kinetic wind scheme, chemical enrichment, super-massive black hole growth, and feedback from active galactic nuclei (Vogelsberger et al. 2013).

Merger trees were constructed from the main \textsc{Illustris-1} simulation using the SUBLINK algorithm, which identifies a unique sub-halo descendant from the next snapshot using a merit function that takes into account the binding energy rank of each particle to discriminate between the potential sub-halo candidates (see Rodriguez-Gomez et al. 2015 for more details on the creation of merger tree of sub-halos and galaxies). The \textsc{Illustris-1} simulation has already been used in previous works related to galaxy mergers. These analyses suggest that  major pair fractions change little or decrease with increasing redshift for $z>1$ (Snyder et al. 2017), which is in agreement with recent surveys, and that  50(20)\% of the ex-situ stellar mass in nearby of  elliptical galaxies comes from major(minor) mergers (Rodriguez-Gomez et al. 2016).

\begin{figure*}
 \begin{subfigure}{1\textwidth}
 \centering
 \includegraphics[width=1\linewidth]{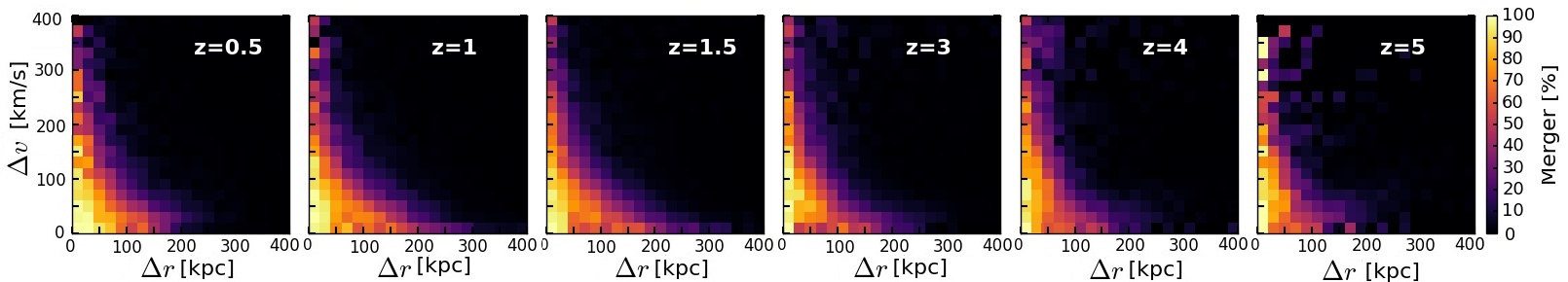}
 \caption{ }
 \label{p1}
 \end{subfigure}
 \begin{subfigure}{1\textwidth}
 \centering
 \includegraphics[width=1\linewidth]{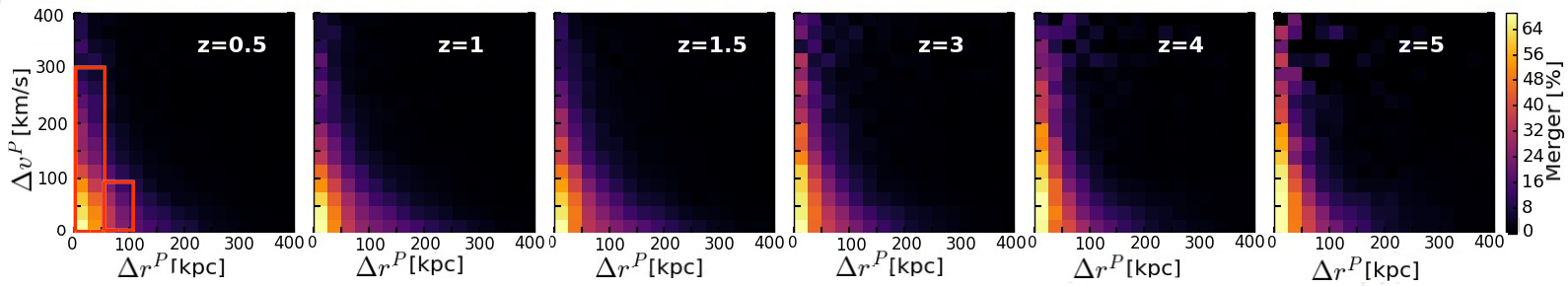}
 \caption{}
 \label{p2}
 \end{subfigure} 
   \caption{(a)\, The galaxy pair velocity-separation distance diagram from \textsc{Illustris-1} simulation for six snapshots at different redshifts, color-coded with respect to the fraction of future mergers within the pair sample. (b)\, Same diagram as (a) but with {\it projected} velocity-separation distances. The two red boxes correspond to the new criteria introduced in sect \ref{new criteria}.}
 \label{simu1}
\end{figure*}

\begin{figure*}[t]
     \begin{tabular}{c}     
       \includegraphics[width=1\textwidth]{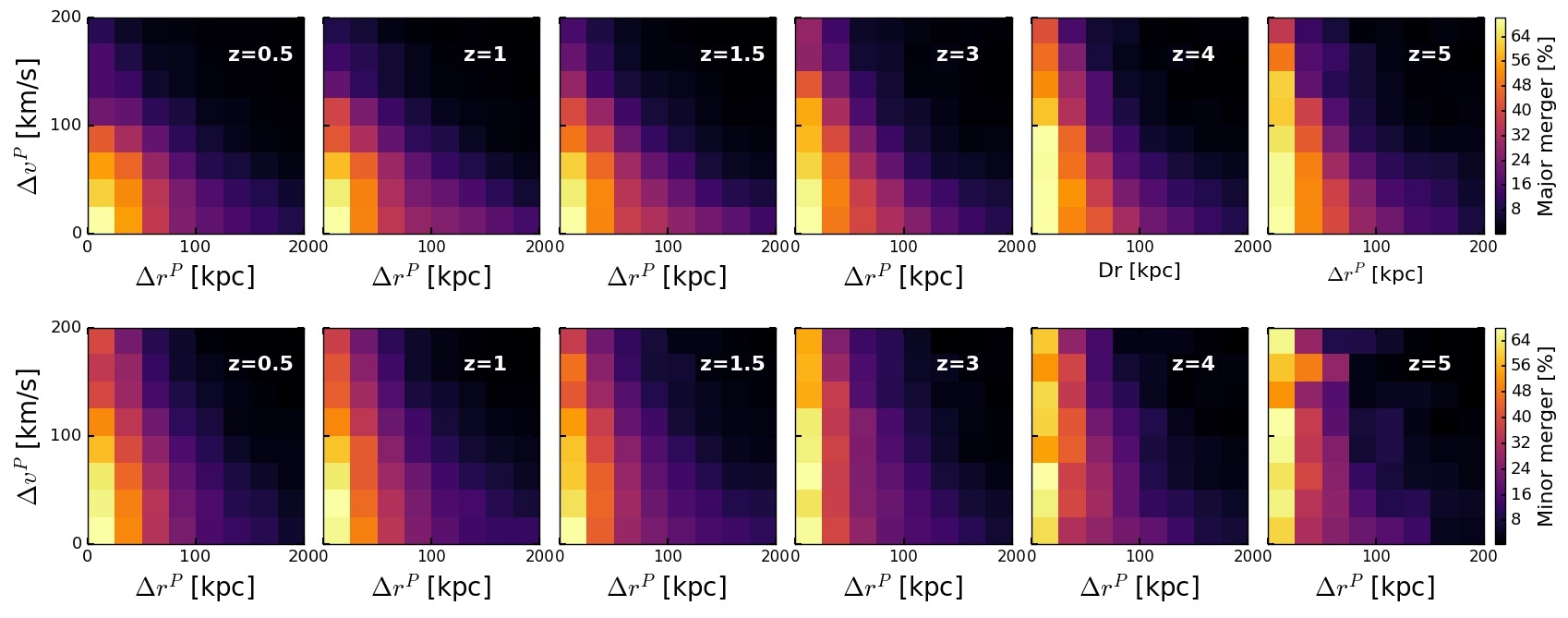}
        \end{tabular}
           \caption{Influence of the galaxy mass ratio in the pair on the velocity-separation distance diagram (zoomed into a box of 200 kpc $\times$ 200 \kms) and the probability of the pair to merge for different redshifts. {\it Top}: Major merger distribution, with a mass ratio between the primary galaxy and its companion within 1:1 and 1:6. {\it Bottom}: Minor merger distribution with a mass ratio in the pair lower than 1:6.}
             \label{simu2_1}
   \end{figure*}

For this analysis, mock catalogs were created from six snapshots of the \textsc{Illustris-1} simulation, corresponding to six different redshifts: $z=$ 0.5, 1, 1.5, 3, 4 and 5. For each of them, merger tree information was generated. The simulation produces galaxies spread over a large range of stellar masses, $10^8 \leq $ \Mstar $\leq 10^{11.5}$\Msun\ (see Fig.\,\ref{fig:massdistrib_illustris}). The lower mass cut of $10^8$\Msun\ is about two order of magnitude higher than the nominal baryonic mass resolution of the \textsc{Illustris-1} simulation. The number of galaxies in these mocks decreases with redshift, from $\sim 75\,000$ at $z=0.5$ to $\sim 9\,000$ at $z=5$. This variation in the number of galaxies is reflected in Figs.\,\ref{simu1} and \ref{simu2_1}, where the statistics decreases at high redshift. Since mock data are versions of the real simulation in which the geometry and selection effects of observational surveys are reproduced, it can  be analyzed using similar methods, which is a powerful advantage for comparisons between theory and observations. 

\section{New criteria for the selection of galaxy close pairs}
\label{criteria}

\begin{figure*}[t]
 \centering
     \begin{tabular}{c}  
       \includegraphics[width=0.65\linewidth]{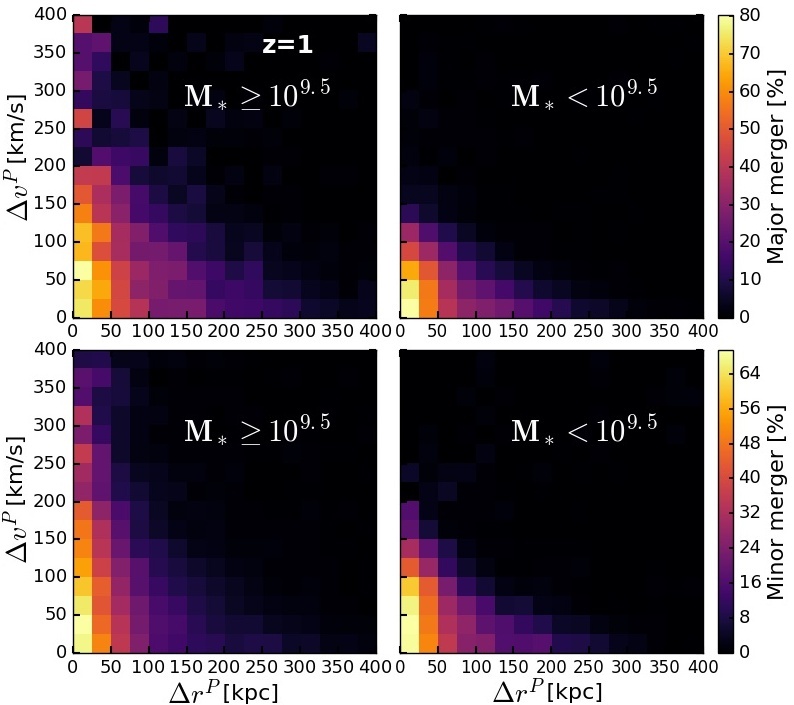}
        \end{tabular}
           \caption{ Influence of the primary galaxy's stellar mass on the merger velocity-separation distance diagram. A constant stellar mass limit of $10^{9.5}$\Msun\ is chosen to distinguish low-mass from massive primary galaxies within the major  ({\it top}) and the minor ({\it bottom}) pair samples at $z = 1$.}
             \label{simu2_2}
   \end{figure*}
%\begin{figure*}
% \begin{subfigure}{1\textwidth}
% \centering
% \includegraphics[width=1\linewidth]{Proj_simus_Maj_Min_ref.jpg}
% \caption{ }
% \label{p1}
% \end{subfigure}
% \begin{subfigure}{1\textwidth}
% \centering
% \includegraphics[width=0.6\linewidth]{Proj_simus_Major_Minor_sep_masse_ref.jpg}
% \caption{}
% \label{p2}
% \end{subfigure} 
%   \caption{(a)\,}
% \label{simu2}
%\end{figure*}

A close pair of galaxies is defined as two galaxies with a small rest-frame relative velocity and projected separation distance in the sky plane. These selection parameters are respectively computed as follows in most observational surveys:
\begin{equation}
r_p^{min} \leqslant r_p = \theta \times d_A (z_m) \leqslant r_p^{max}, 
\end{equation}
where $\theta$ is the angular distance between the two galaxies, $d_A (z_m)$ is the angular scale (in kpc arcsec$^{-1}$) and $z_m$ is the mean redshift of the two galaxies, and:
\begin{equation}
\Delta v = \frac{c \times |z_1-z_2|}{(1+z_m)}\leqslant \Delta v_{max}, 
\end{equation}
where $z_1$ and $z_2$ are the redshifts of each galaxy in the pair and $c$ is the speed of light.

For each of these two criteria, a wide range of values can be found in the literature (see sect.\ref{sec:intro}). %$r_p^{min} \sim 0-10$ h$^{-1}$kpc, $r_p^{max} \sim 20-50$ h$^{-1}$kpc and $\Delta v_{max} \sim %500$ \kms, which can partially explains the large scatter between measurements. These criteria %are expected to offer a good compromise between chance pairing contamination and pair %statistics, with galaxies having a high probability of merging since $30$ h$^{-1}$kpc  appears %to be the approximate scale on which the majority of the  pairs start to exhibit interacting %features such as tidal tails, distortions or SFR enhancements (Patton et al. 2000;  Alonso et %al. 2004; Nikolic et al. 2004; Patton \& Atfield 2008).
%However, recent works based on the SDSS have shown that effect of galaxy interactions can also be detected in galaxy pairs with separation greater than 50 h$^{-1}$kpc, with SFR enhancements found at projected separations up to 150 h$^{-1}$kpc (Scudder et al. 2012; Patton et al. 2013).
%We also have to take into account that the close pair method does not precisely trace merger event, but rather looks for likely potential future mergers since some close pairs will not finally merge.
In the following subsections, we try to improve these parameters by analyzing the relation between the velocity-distance relative separation of a close pair of galaxies and the probability that this pair will merge in the future.

\subsection{Galaxy pairs identification}
For each of the six mock catalogs created from \textsc{Illustris-1} simulation (see section\,\ref{Illustris_data}), we applied selection techniques that are commonly used in observational surveys. Knowing the position and velocity of each galaxy in real space, we detect pairs of galaxies with a difference in relative velocity amplitude $\Delta v \leqslant 500$  \kms, since most studies have shown  that  pairs
 with $\Delta v > 500$  \kms\ are not likely to be gravitationally bound (Patton et al. 2000; De Propris et al. 2007), and a separation distance, $\Delta r \leqslant 500$  kpc, which allows us to explore a large range of values for the separation distance criterion.
From the merger trees, we can then follow the descendant branch informations for each sub-halo in the subsequent snapshot, and so on, until $z=0$ and thus identify which ones of the galaxy pairs become a true merger in the future. 

\subsection{Results}
\subsubsection{Projection effect} 
Figure\,\ref{simu1}(a) shows the relation between the true velocity-distance separation of the galaxy pairs in real space and their probability to merge for the different redshifts. As expected the probability of a pair to merge decreases both with the separation distance, $\Delta r$, and with the velocity difference, $\Delta v $, of the galaxy pair. The decrease is slower for $\Delta v $ than for $\Delta r$. Thus, a galaxy pair within a separation distance of $\Delta r\leqslant  50 $ kpc and velocity difference of $\Delta v \leqslant 200$  \kms\  has between 100 and 80\% of chance to merge by $z=0$. For the highest redshift snapshots, $z=4$ and $5$, statistical effects, due to much lower numbers of pairs, begin to appear for $\Delta v \geqslant 400$  \kms.

However true, unprojected, velocity differences and physical separation distances are not available in observations. Thereby we use projected values of the relative velocity and distance which reflect the projected separation distance in the sky plane and the rest-frame velocity difference in redshift space from observations.
The probability for a galaxy pair to merge as a function of its position in the projected velocity-distance diagram ($\Delta v^{P}$ vs.\, $\Delta r^{P}$) is shown  in Figure\,\ref{simu1}(b). 
Using projected values clearly affects the probability of a pair to merge because of contamination effects, dropping the probability to 70\% for a pair with $\Delta r^{P} \leqslant 25$ kpc and $\Delta v^{P} \leqslant 100$ \kms. In the projected space, pairs with $\Delta v^{P} \geqslant 300$ \kms\ have less than 10\% chance to end up as a merging system.

\subsubsection{Dependence on stellar mass }
Interacting galaxies can end up in a merging system if the two colliding galaxies do not have enough momentum to overcome the gravitational hold they have on one another and continue their courses after the collision. Velocities, angles of the collision, sizes, relative composition or masses are all parameters that can affect the result of two colliding galaxies. The more massive the primary galaxy is, the more gravitational pull it will have, the harder it will be for its companion to liberate itself from its hold. 
%Likewise for two interacting galaxies with a very low mass ratio, the stellar mass difference between the two galaxies is such that it is very unlikely the satellite galaxy will be able to escape the gravitational field of the primary one.

An attempt to study the influence of the mass ratio on the relation between the velocity-separation distance diagram of the galaxy pairs and their probability to merge is shown in Fig.\,\ref{simu2_1}. The influence of the primary galaxy stellar mass on this relation is shown in Fig.\,\ref{simu2_2}.

First, we use the mass ratio between the two galaxies to discriminate the pairs. Figure \ref{simu2_1} shows the relation between the projected velocity-distance separation of the pair and their probability to merge for major close pairs, with a stellar mass ratio higher than 1:6 (as adopted in Ventou et al. 2017), and minor close pairs, i.e.\, with stellar mass ratio lower than 1:6. The main difference seen in Fig.\,\ref{simu2_1} between the two samples comes from the rest-frame relative velocity condition. For a fixed projected separation distance $\Delta r^{P} \leqslant 25 $ kpc, $\sim$70\% of the major close pairs will merge if their rest-frame relative velocity $\Delta v^{P}$ is lower than $\sim$ 50 \kms (see Fig.\ref{simu2_1}, top panels), whereas the same fraction of mergers will be reached by minor close pairs with $\Delta v^{P} $ up to 100 \kms (see Fig.\ref{simu2_1}, bottom panels).

We further separate our sample into two regimes using the stellar mass of the primary galaxy, i.e.\, the most massive one of the pair, as a limit. In Fig.\, \ref{simu2_2} we distinguish, for the $z=1$ snapshot, between low-mass and massive galaxies within the major and minor close pair samples by applying a separation limit of $10^{9.5}$\Msun, similar to the limit adopted in Ventou et al. (2017).

As for the major-minor discrimination, the stellar mass separation affects mainly the condition on the rest-frame velocity.
For a primary galaxy with a stellar mass, \Mstar\ $> 10^{9.5}$\Msun, a pair within $\Delta r^{P} \leqslant 25 $ kpc and  $\Delta v^{P} \leqslant 150 $  \kms\ has between 75 and 60\% chance to merge, for a major and minor close pair respectively. However, for pairs with a lower-mass primary galaxy, \Mstar\ $\leqslant 10^{9.5}$\Msun, and for the same probability to merge, the threshold in relative velocity is smaller: $\Delta r^{P} \leqslant 25 $ kpc and  $\Delta v^{P} \leqslant 75 $  \kms. Similar results are obtained for the other redshift snapshots, showing that these conditions have almost no dependance with redshift.

To summarize, the stellar mass of the galaxies involved in the pair will mostly have an impact on the rest-frame relative velocity selection criterion. Massive primary galaxies with their strong gravitational pull can retain satellite galaxies with larger relative velocity difference than lower-mass galaxies.

\subsubsection{New criteria for pair selection and weighting scheme} 
\label{new criteria}

\begin{figure}[t]
	\includegraphics[width=\columnwidth]{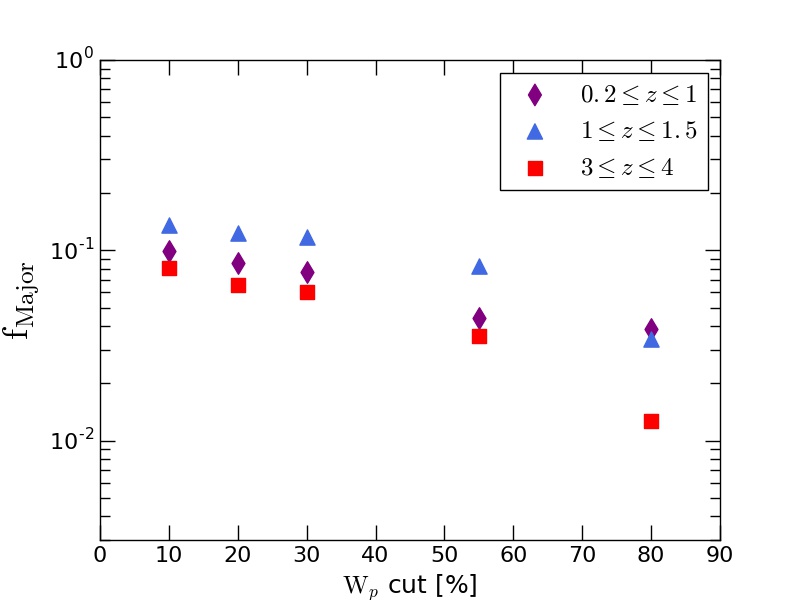}
   \caption{Major merger fraction of close pairs selected in MUSE deep observations over the Hubble Ultra Deep Field (so-called UDF-mosaic) for three redshift bins and five values of merging probability threshold (10, 20, 30, 55, and 80\%). The choice of a 30\% threshold is motivated by the convergence of the merger fraction below this value.}
   
	\label{fig:testcut}
\end{figure}

From this analysis we propose new criteria for the selection of galaxy close pairs.  We define a close pair as two galaxies within a limited projected separation distance in the sky plane and a rest-frame relative velocity of:
$$\begin{cases}  \ 5 \leqslant \Delta r^P \leqslant  50 \ {\rm kpc} \ \text{  and  } \ \Delta v^P \leqslant  300  \ {\rm km\ s^{-1}} \\
 {\rm or} \ 50 \leqslant \Delta r^P \leqslant 100 \ {\rm kpc} \ \text{ and } \ \Delta v^P \leqslant 100 \ {\rm km\ s^{-1}}\\
\end{cases}$$
With these parameters, all close pairs with at least 30\% of chance of merging are considered, regardless of the mass ratio or stellar mass of the primary galaxy (see red boxes in lower left panel of Fig.\,\ref{simu1}) . We choose this threshold of 30\% as the fraction of mergers converge below this value. This is illustrated in Fig.~\ref{fig:testcut} where we show the major merger fraction computed in the Hubble Ultra Deep Field for three redshift intervals as a function of five different values of merging probability threshold (10, 20, 30, 55, and 80\%). Adding pairs with a probability to merge below the 30\% threshold would increase the major merger fraction by a few percent only.

The limit $\Delta r^P_{min} \sim 5$ kpc comes from the limitation in spatial resolution of the MUSE data (see Ventou et al. 2017). Indeed, two galaxies within an angular separation of  $\theta \leqslant 0.7\arcsec$ which corresponds approximately to $\Delta r^P_{min}\sim 5$ h$^{-1}$kpc at $z\sim 1$, are nearly impossible to distinguish and would appear as a blended object. Further in the analysis, a corrective term is applied to the expression of the merger fraction to account for the missing pairs.
We note that these values are similar to those applied by Scudder et al. (2012) and  Patton et al. (2013) in their SDSS-based study of the SFR enhancement in pairs of interacting galaxies.

Based on a least-squares fit to the simulated datasets shown in Fig.\,\ref{simu1}(b) with a non-linear regression, a new weighting scheme can be applied to the merger fraction which takes into account the probability of the galaxy pair to merge as a function of their relative velocity (in kpc) and projected separation (in \kms) distances (see Appendix~\ref{reg}  for more details):
\begin{equation}
W(\Delta r^P ,  \Delta v^P)= 1.407\pm0.035 \ e^{-0.017\pm0.0004 \ \Delta r^P \ -0.005\pm0.0001 \ \Delta v^P} 
\label{eq:weight}
\end{equation}
 If we further divide the sample by the stellar mass of the primary galaxy (see Fig.\,\ref{simu2_2}), we propose the following two equations for  galaxies with a stellar mass above or below \Mstar$=10^{9.5}$\Msun respectively:

\begin{equation}
$$W(\Delta r^P ,  \Delta v^P)= 
\begin{cases}1.617\pm0.064 \ e^{-0.016\pm0.0006 \ \Delta r^P \ -0.008\pm0.0003 \ \Delta v^P} \\
1.375\pm0.052 \ e^{-0.018\pm0.0005 \ \Delta r^P \ -0.004\pm 0.0002 \ \Delta v^P} \\
\end{cases}$$
\label{eq:weightbymass}
\end{equation}

%\begin{equation}
%W(r_p ,  \Delta v)= 1.617\pm0.064 \ e^{-0.016\pm0.0006 \ r_p \ -0.008\pm0.0003 \ \Delta v} 
%\end{equation}
%{{\bf The one standard deviation errors on the parameters are respectively: 0.0641 ,0.0006 and 0.0003. }}
%{{\bf Then  $log(M) \leq 9.5$}}
%\begin{equation}
%W(r_p ,  \Delta v)= 1.375\pm0.052 \ e^{-0.018\pm0.0005 \ r_p \ -0.004\pm 0.0002 \ \Delta v} 
%\end{equation}
%{{\bf The one standard deviation errors on the parameters are respectively: 0.0518 ,0.0005 and 0.0002. }}
This new corrective term allows us to give an estimate of the close pair fraction that reflects more accurately the true merger fraction. The following sections present the application of these new selection criteria and weighting scheme on MUSE deep fields in order to derive the evolution of the major and minor galaxy pair fractions over the last 13 Gyr.

\section{Data description}

\begin{figure}[t]
	\includegraphics[width=\columnwidth]{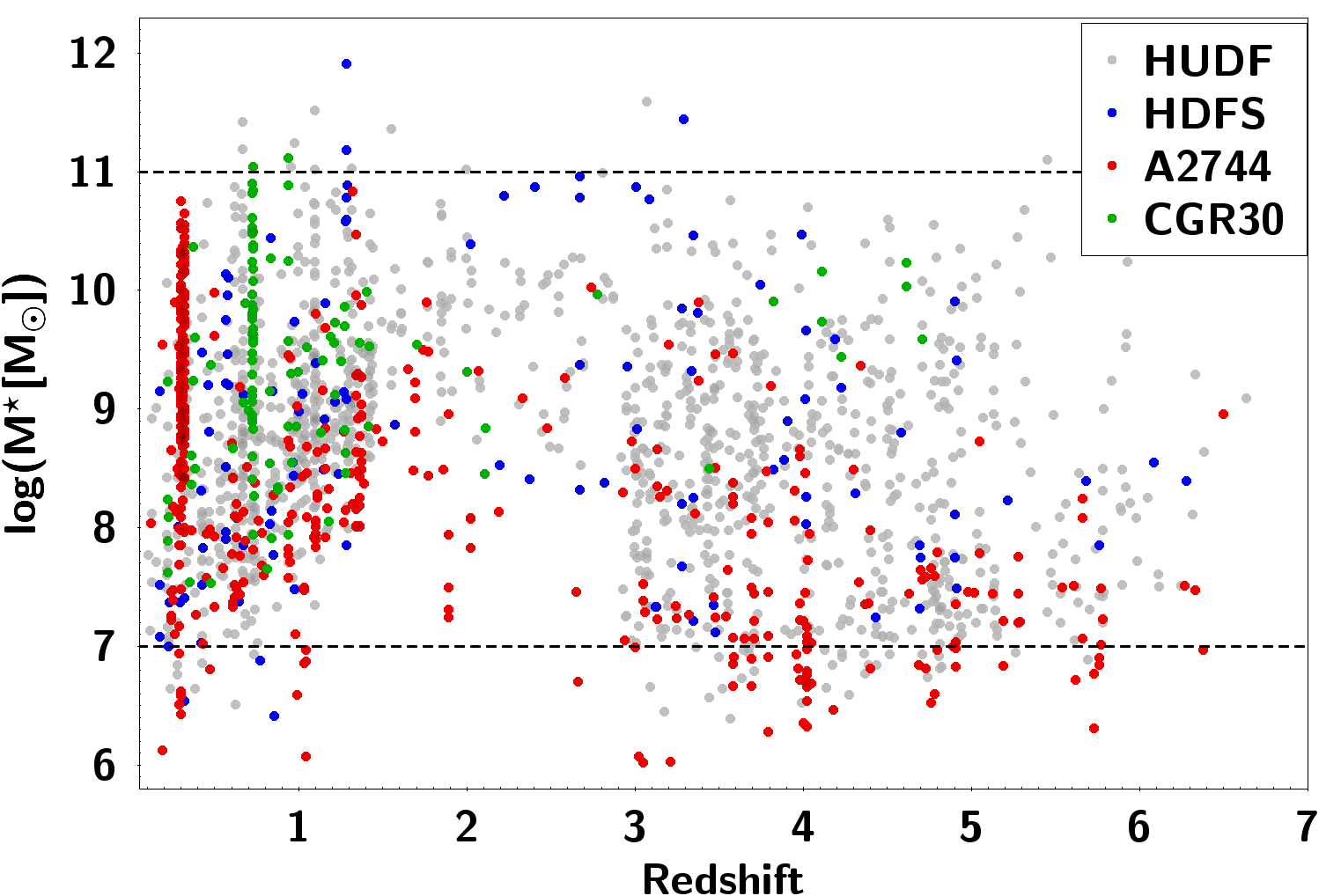}
   \caption{Stellar mass as a function of redshift for the parent sample of 2483 galaxies drawn from MUSE deep observations in HUDF (grey dots), HDFS (blue dots), A2744 (red dots), and CGR30 (green dots). Thanks to the high sensitivity of MUSE, star-forming galaxies are identified with a secure spectroscopic redshift up to $z\sim 6$ over a large range of stellar masses ($\sim 10^7-10^{11}$\Msun), except in the so-called ``redshift desert'' ($z\sim 1.5-2.8$) where galaxies in the low-mass regime (\Mstar\ $\sim 10^7-10^8$ \Msun) are detected behind the A2744 lensing cluster only.}
	\label{fig:mstarvsz}
\end{figure}

 \begin{figure}[t]
	\includegraphics[width=\columnwidth]{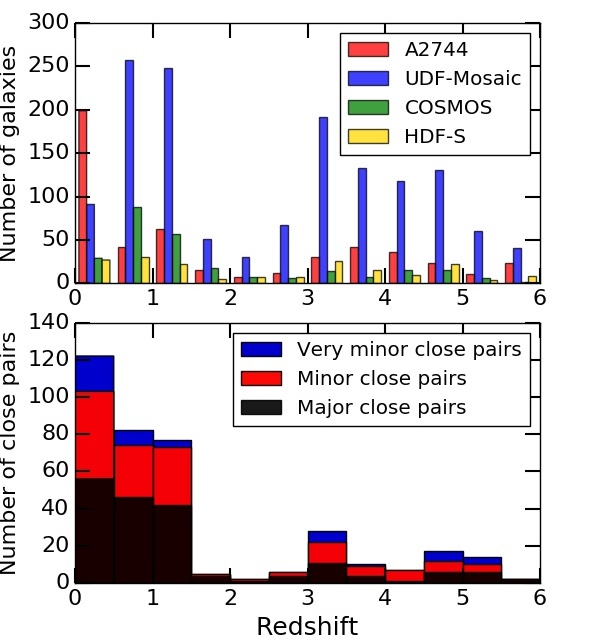}
   \caption{{\it Top}: The spectroscopic redshift distribution of the parent galaxies in the four MUSE data cubes used in this analysis. {\it Bottom}: Redshift histogram of the close pairs sample showing the contribution of major (black), minor (red), and very minor close pairs (blue).}
	\label{fig:zdistrib}
\end{figure}

The analysis presented in this paper is based on MUSE (Multi Unit Spectroscopic Explorer, Bacon et al.\,2010) observations obtained during the last commissioning run of the instrument in August 2014 and 1.5 years of MUSE Guaranteed Time Observations (GTO), from September 2014 to February 2016.

\subsection{Parent sample}
\label{sample}

\begin{figure*}[!h]
 \centering
 \includegraphics[width=1\linewidth]{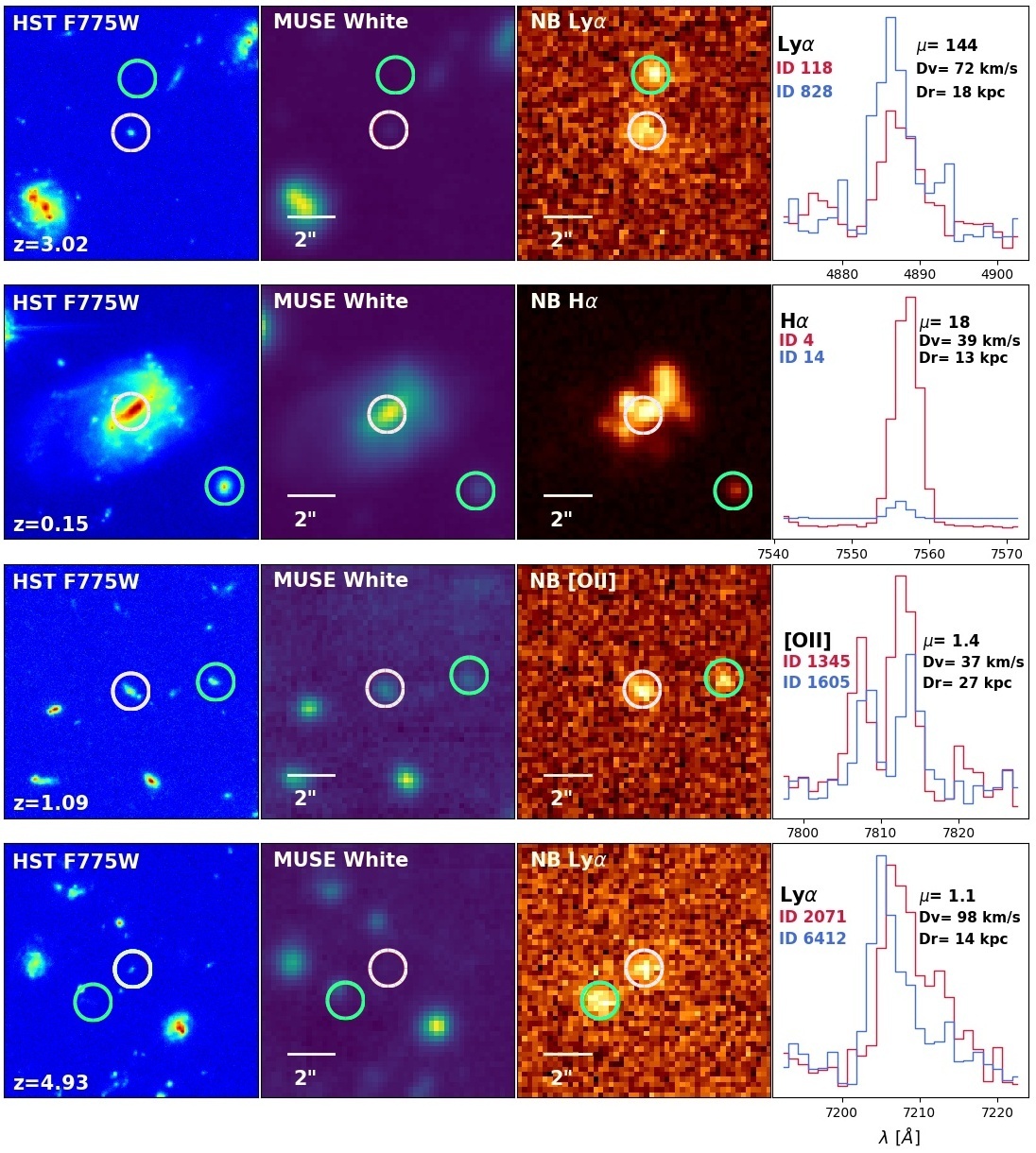}
 \caption{Examples of close pairs of galaxies, at different redshifts and with various mass ratios in the UDF-Mosaic field. {\it From left to right}: HST image in the F775W filter with the redshift of the primary galaxy, MUSE reconstructed white light image, narrow-band image of one of the brightest emission lines of the pair, and the zoomed spectra of the two galaxies around this line (red for the primary galaxy and blue for its companion) with the labeled MUSE ID, mass ratio, relative velocity difference and projected separation distance. Images are $10$\arcsec\ in linear size and centered around the primary galaxy, i.e.\, the most massive one, circled in white. The green circle indicates the location of its companion.
The first two close pairs correspond to the very minor and minor regimes, the last two are major close pairs at low and high redshift.
}
 \label{im_pair}
\end{figure*}
For this analysis, a large spectroscopic sample of 2483 galaxies is constructed from MUSE deep observations over four different regions of the southern sky, the Hubble Deep Field South (HDF-S; Bacon et al., 2015), the Hubble Ultra Deep Field (HUDF; Bacon et al., 2017), the lensing cluster Abell 2744 (Mahler et al., 2018), and the galaxy group COSMOS-Gr30 at $z\sim 0.7$ (Epinat et al., 2018). The two first MUSE datasets in HDFS and HUDF are already described in Ventou et al.\,(2017). In the following subsections, we describe the additional MUSE datasets in Abell 2744 and COSMOS-Gr30.

\subsubsection{Abell 2744}
Abell 2744 was observed as part of a GTO program aimed at probing the highly magnified regions of massive lensing clusters (PI: J.\,Richard). The resulting data cube is a $2\arcmin \times 2\arcmin$ mosaic centered around $\alpha=00h 14' 20.95"$ and $\delta=-30^o 23' 53.88"$ with  exposure times ranging from 4h to 7h. Instrumental setup was similar to HUDF observations. Sources were extracted using three complementary detection methods described in Mahler et al. (2018): spectral extraction at the location of known faint sources in the deep Hubble Frontier Field images, emission line detection based on narrow-band filtering in the MUSE data cube using the software MUSELET\footnote{MUSELET is an analysis software released by the consortium as part of the MPDAF suite http://mpdaf.readthedocs.io/en/latest/muselet.html}, and finally manual extraction for sources found by visual inspection and not detected with the previous methods. Overall the spectroscopic redshift of 514 sources was measured, with 414 new identifications (Mahler et al., 2018). For this study, we kept one galaxy only for all the confirmed multiple-images systems. The source positions were corrected for lensing effects and estimated in the source plane (Mahler et al., 2018). Howerver, lensing does not affect the redshift and velocity differences measured in the MUSE data.

\subsubsection{COSMOS-Gr30}
The deep observations of the galaxy group COSMOS-Gr30 at $z\sim 0.7$ are part of a large GTO program that aims to study how the environment affect galaxy evolution over the past 8 Gyr (PI: T. Contini). A single field of $1\times 1$ arcmin$^{2}$ and 10h exposure time was obtained, comprising 40 exposures of 900 seconds. 
%with 90 degrees rotation between exposures. 
The data cube presents the same spatial and spectral sampling characteristics as for the HUDF and Abell 2744. The seeing was estimated to be around $0.68$\arcsec\ at $7000$ \AA\ (Epinat et al., 2018). 
As for the UDF-Mosaic, sources were selected from the COSMOS2015 photometric catalogue (Laigle et al., 2016), complemented by emission-line detection using ORIGIN software (Bacon et al. 2017). A customized version of the redshift finding code MARZ (Hinton et al., 2016) was used to assess the spectroscopic redshift of the sources. The final catalog consists of 208 spectroscopic redshifts.

\subsubsection{Redshift and stellar masses}
\label{sec:zandmass}
By combining the catalogs associated with each of the four surveys, we build a parent sample of 2483 galaxies with spectroscopic redshift up to $z\sim 6$. Each redshift measurement is assigned a confidence level based mostly on the detected spectral features (see details in Inami et al.\,2017). Confidence 3 and 2 are secure redshifts based on several spectral features, such as strong emission lines, or a clearly identified single one (mainly \oii\ and \lya), or strong absorption features. Confidence 1 is a tentative redshift with uncertainties on the nature of the feature from the line profile, most of the time \lya\ versus \oii. This redshift confidence is later taken into account in the merger fraction estimate, a weight is thus applied to distinguish between secure galaxy pairs involving two galaxies with a confidence level of 3 or 2, and unsecure ones involving at least one galaxy with confidence 1. As in previous paper (Ventou et al. 2017), we used the empirical relation between the velocity shift of the \lya\ emission peak relative to the systemic velocity and the FWHM of the line (Verhamme et al. 2018) to compute the systemic redshift of all the \lya\ emitters. 

Stellar masses were derived using FAST (Fitting and Assessment of Synthetic Templates), a code that fits stellar population synthesis templates to broad-band photometry and spectra (Kriek et al.\,2009). We assume for all four fields a Chabrier (2003) initial mass function, an exponentially declining star formation history and the dust attenuation law from Calzetti et al.\,(2000). As described in Ventou et al.\,(2017), extended UV-to-NIR ACS and WFC3 photometric measurements (Rafelski et al.\, 2015) were used for the UDF-Mosaic, with the addition of mid-infrared IRAC photometry from the {\it GOODS Re-ionization Era wide-Area Treasury from Spitzer} program to better constrain the stellar mass of high-redshift galaxies ($z\geqslant 3$). The optical-NIR photometric bands used for the HDF-S are also listed in Ventou et al.\,(2017). 
As described in Epinat et al.\,(2018), the photometric measurements for COSMOS-Gr30 come from the extensive dataset available in the COSMOS field (Scoville et al. 2007) and summarized in the COSMOS2015 catalog (Laigle et al., 2016). It includes infrared and far-infrared  photometry from Spitzer and Herschel, radio data from the VLA, UV-to-infrared from HST-ACS, SDSS, VIRCAM/VISTA camera, WIRCam/CFHT and  MegaCam/CFHT camera as well as HSC/Subaru $Y$ band and SuprimeCam/Subaru, near and far ultraviolet measurements from the Galaxy Evolution Explorer, Chandra and XMM observations for X-ray data (see Epinat et al., 2018 for details). For the lensing cluster A2744, seven HST bands (ACS; F435W, F606W, F814W and WFC3; F105W,
F125W, F140W, F160W) were used. A median boxcar subtraction was applied to these images in order to better estimate the stellar masses of faint background galaxies, which can be contaminated by the light of the cluster galaxies. Results are all corrected for lensing magnification effects.

The final parent galaxy sample assembled from the four MUSE deep surveys probes a large domain in stellar mass, from $\sim10^{7}-10^{11}$ \Msun, distributed over a large redshift range $0.2 \leqslant z \leqslant 6.8$ (see Fig.\,\ref{fig:mstarvsz}). Note that no stellar mass has been derived for the few \lya\ emitters in the UDF-Mosaic which are not detected in deep HST images (see Inami et al. 2017). These galaxies are not considered further in the analysis.

Figure\,\ref{fig:zdistrib} ({\it top}) shows the spectroscopic redshift distribution of the parent galaxy sample for all individual fields. 
Peaks in the histograms account for particular structures detected in each data cube. In the UDF-Mosaic an over-dense structure is detected around $z\approx 1$ (see Inami et al.\, 2017), the peak at $z\approx 0.3$ in A2744 corresponds to the galaxy cluster, overall the redshifts of 156 cluster members were measured from MUSE observations over this region (Mahler et al., 2018). The green peak around $z\approx 0.7$ represents the galaxy group Gr30 in COSMOS.
The dearth of spectroscopic redshifts between $1.5\leqslant z\leqslant 2.8$ is expected, as it covers the well known "redshift desert" interval for optical instrument such as MUSE. Due to the absence of bright emission lines in this range (in between \lya\ and \oii), the redshifts are measured mainly on absorption features or \ciii\ emission-line doublet.

\subsection{Close pair sample}

 \begin{figure}[t]
	\includegraphics[width=\columnwidth]{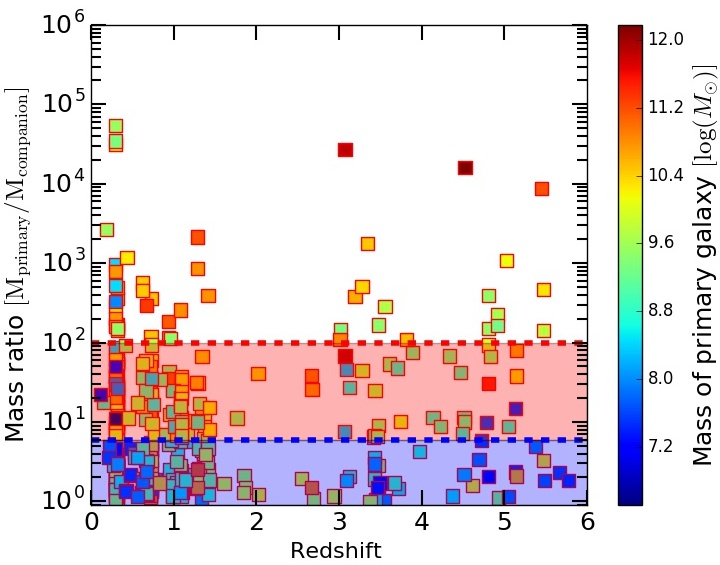}
   \caption{ Stellar mass ratio and redshift distribution of the whole close pairs sample from the combined analysis of the four MUSE deep fields. Symbols are color-coded with respect to the primary galaxy's stellar mass, ie.\,the most massive one in the pair. The dashed lines indicate the mass ratio (primary over companion galaxy) limits chosen to distinguish major close pairs (limit of 6, blue dashed line and colored area) from minor (between a mass ratio limit of 6 and 100, red dashed line and colored area) and very minor ones (mass ratio greater than 100).}
	\label{fig:Mass}
\end{figure}
 \begin{figure}[t]
	\includegraphics[width=\columnwidth]{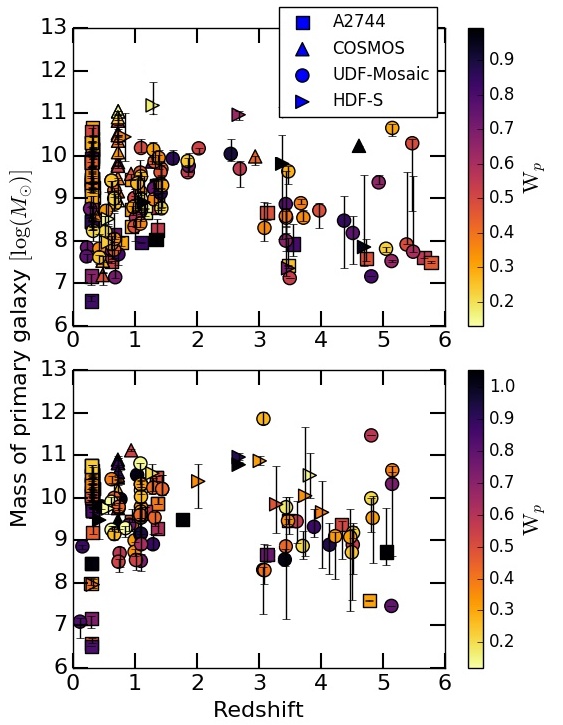}
   \caption{Stellar mass of the primary galaxy as a function of redshift for our major ({\it top}) and minor ({\it bottom}) close pair samples. Right pointing triangles correspond to close pairs in the HDF-S, other triangles to close pairs in the COSMOS-Gr30 field, squares are pairs from A2744, and circles from the UDF-Mosaic. Symbols are color-coded with respect to the weight, $W_{p}$ described in section \ref{new criteria}, used for the computation of merger fractions. Darker pairs have a higher probability to merger by $z=0$ and will thus have a higher contribution on the estimated merger fraction. Except in the redshift ``desert'' ($z\sim 1.5-2.8$), we have a fairly good stellar mass completeness level for major close pairs down to a primary galaxy stellar mass of $\approx 10^7$\Msun. However, for the minor close pairs sample, a lower mass limit of $10^9$\Msun\ must be applied to keep a reasonable level of completeness for the merger fraction estimates.}
	\label{fig:wp}
\end{figure}

Applying the criteria defined in section \ref{new criteria} to the MUSE data set, a total of 366 close pairs of galaxies were identified. About 44\% of them were detected in the UDF-Mosaic, 40\% in A2744, 9\% in COSMOS-Gr30, and 7\% in HDF-S. As mentioned  in sect.\,\ref{sec:zandmass}, we do not include in this sample the few $z > 3$ pairs detected in the UDF-Mosaic ($\sim 1.8$\% of the total pair sample) of one or two \lya\ emitters without a HST counterpart, and thus without any stellar mass estimate. 

The mass ratio, defined as the ratio between the stellar mass of the companion and that of the primary galaxies, is used as a proxy to divide this sample into major mergers, with a mass ratio of 1:1$-$1:6 as chosen in Ventou et al.\,(2017), minor mergers (1:6$-$1:100), and very minor mergers with a mass ratio lower than 1:100. In this last regime the primary galaxy is so much more massive than its companion that it is getting closer to the regime of smooth gas accretion than to a galaxy merger. The secondary galaxy is completely stripped and absorbed by the massive one. 

Within our sample, we thus identify a total of 179 major close pairs, 140 minor and 47 very minor close pairs, distributed over a broad range of redshift $0.2 \leqslant z\leqslant 6$ (see Fig.\,\ref{fig:zdistrib}, bottom). 
As stated before, the peaks around $z\approx 0.3 $ and $z\approx 0.7$ in the redshift distributions are due to the lensing cluster and galaxy group respectively. More than 30(32) major(minor) close pairs are detected at high redshift $z \geqslant 3$. Examples of close pairs in each mass ratio regimes are displayed in Fig.\,\ref{im_pair}. The first two raws correspond to a very minor and minor pair at $z \sim 3$ and $z \sim 0.15$ respectively. The last two rows are both major close pairs at $z\sim 1$ and $z\sim 5$. The catalogs of major and minor pairs are listed in Tables\,\ref{table:majorpairs} and \ref{table:minorpairs}, respectively.

Figures\,\ref{fig:Mass} and \ref{fig:wp} reveal the mass ratio and stellar mass domain of our samples. With deep enough MUSE observations, we manage to probe galaxy pairs with very low mass ratios (down to 1:10$^4$, see Fig.\,\ref{fig:Mass}) at any redshift, except for the "redshift desert" interval. Pairs with a mass ratio lower than 1:100 are considered as very minor, close to the smooth accretion regime. Likewise the stellar masses range of the primary galaxies extend over 4 dex, from $\sim10^{7}$ to $10^{11}$\Msun\ (see Figs.\,\ref{fig:mstarvsz} and \ref{fig:wp}). Within this mass range, the major close pair sample has a good level of stellar mass completeness, as already discussed in Ventou et al.\,(2017). Three of the four MUSE datasets used in this analysis (HUDF, A2744 and CGR30) have a similar exposure time of 10 hours, leading to a homogeneous stellar mass completeness down to $\sim10^7$\Msun\ up to redshift z$\sim6$. However, as can be seen in Fig.\,\ref{fig:mstarvsz}, a significant number of galaxies identified in the A2744 field have very low stellar masses thanks to the lensing cluster magnification allowing to unveil star-forming galaxies down to $\sim 10^6$ \Msun. Even if the MUSE observations in the HDFS are significantly deeper (30 hours) than in other fields, we do not reach galaxies with lower stellar masses, still keeping an homogeneous completeness over the four datasets. This is likely due to HDFS data being taken during commissioning when the calibration sequence was not fully established and with a slightly older data reduction procedure. However, due to the extended mass ratio range down to 1:100 considered for the minor close pairs, the low-mass  threshold must be reduced to keep a fairly good mass completeness for this regime. Thereby, for the minor close pair sample, we adopt a low-mass cut of $10^{9}$\Msun. This effect is even more dramatic for very minor pairs. In this regime, the mass completeness is too poor for the sample to be useful. We thus focus the rest of the analysis on the major and minor samples,  within their corresponding mass ranges, and estimate the associated fractions.

In Fig.\,\ref{fig:wp}, the weighting scheme described in equation \ref{eq:weight} is also shown, differentiating between pairs with a high probability to merge (darkest symbols) and the others, which will have a lower contribution to the merger fractions estimated in the next section.

\section{Evolution of the galaxy major and minor merger fraction up to $z\approx 6$ in MUSE deep fields}
\label{merger_fraction}

In order to probe the evolution of the galaxy merger fraction along cosmic time, the redshift range is divided into different bins containing enough pairs to be statistically significant. We thus follow the division adopted in Ventou et al. (2017), with the lowest redshift bin corresponding to the interval $0.2\leqslant z_{r}< 1$, then $1 \leqslant z_{r}< 1.5$ which end up with the loss of the \oii\ emission-line  in the MUSE spectral range, following the "redshift desert" domain $1.5\leqslant z_{r}< 2.8$ and two more bins $2.8\leqslant z_{r}< 4$ and $4\leqslant z_{r}\leqslant 6$ for the highest redshift close pairs.

\begin{figure*}[t]
     \begin{tabular}{c}
     
       \includegraphics[width=1\textwidth]{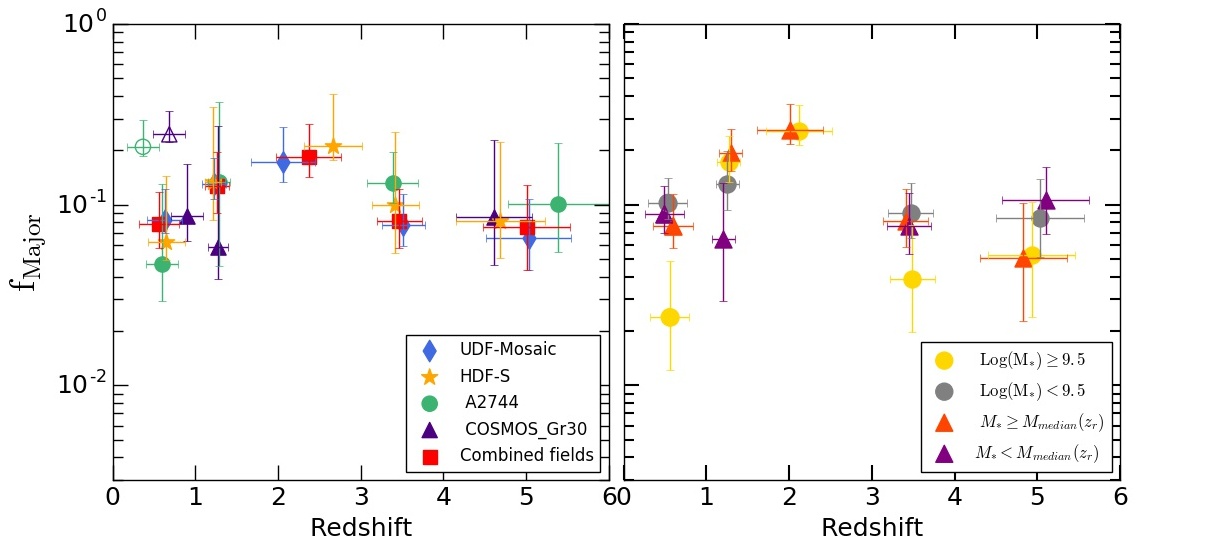}

        \end{tabular}
           \caption{Evolution of the galaxy major merger fraction up to $z\sim 6$ from MUSE deep fields. {\it Left}: Blue diamonds, green circles, yellow stars and purple triangles are estimates of the fraction from the UDF-mosaic, Abell 2744, HDF-S and COSMOS-Gr30 regions respectively, whereas red squares correspond to the fraction for the combined analysis of the MUSE data.  For the lowest redshift bin, fractions were computed without (filled symbols) and with (open symbols) members of the galaxy cluster A2744 and galaxy group COSMOS-Gr30. {\it Right}: Evolution of the major merger fraction for two ranges of stellar mass, assuming first a constant separation limit of \Mstar\ $= 10^{9.5}$\Msun\ (grey and yellow circles show the MUSE estimates for low-mass and massive galaxies respectively), then taking the median mass of the parent sample in each redshift bin as the limit (orange and purple triangles). As for the combined fraction of the left panel, the fractions were computed without taking into account the clusters and group members. The median stellar mass estimated in the range $10^7-10^{11}$\Msun\ for each redshift intervals are listed in Table\,\ref{table:1}.}
             \label{fig:major_frac}
   \end{figure*}

\begin{figure}[t]
 \includegraphics[width=1\columnwidth]{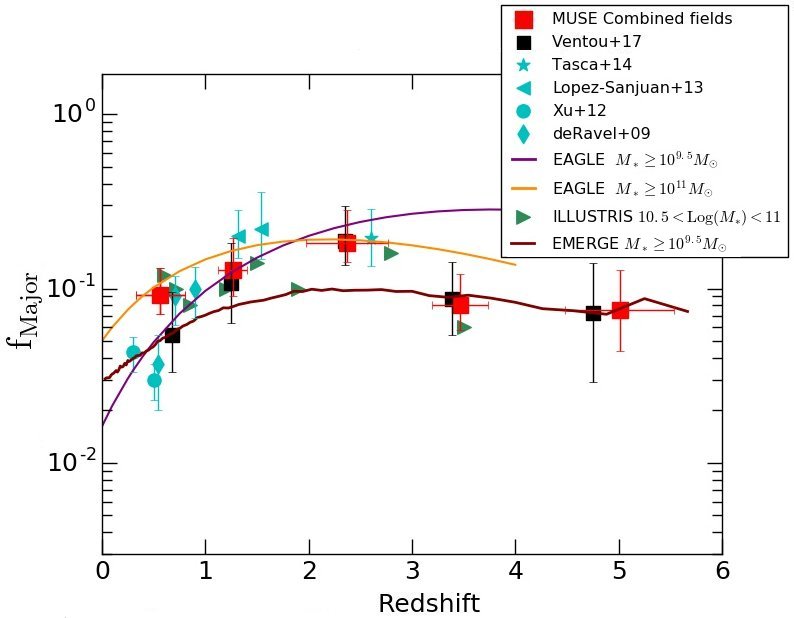}
  \caption{The major merger fraction compared to previous close pairs count studies and recent simulations. Combined major merger fractions from this work (red squares) are compared to previous estimates from MUSE observations (black squares: Ventou et al. 2017) and other surveys (light blue symbols: de Ravel et al. 2009; Xu et al. 2012; Lopez-Sanjuan et al. 2013; Tasca et al. 2014). The purple and orange solid lines show the  predictions from the \textsc{Eagle} simulations for two different mass ranges (see Qu et al.\,2017 for details). Green triangles correspond to the major pair fractions estimated in the \textsc{Illustris} simulation (Snyder et al.\,2017). Finally, the predictions of pair fractions up to $z\sim 6$, with a lower stellar mass cut-off of $10^{9.5}$\Msun, from the \textsc{Emerge} simulations (O'Leary et al. in prep.) are shown with the brown line.}
  \label{fig:major_frac_2}
\end{figure}
   
\subsection{Major merger fraction}
\label{major_merger_fraction}

\begin{table*}
\caption{Major merger fractions up to $z\approx 6$ from MUSE deep observations for different redshift and stellar mass intervals. Cols.\,(1) and (2): Range of the redshift bin and its associated mean redshift for the close pairs sample. Col.\,(3): Median value of stellar mass for the pair sample. Cols.\,(4) and (5): Number of pairs, $N_p$, and galaxies, $N_g$.  Col\,(6): Major merger fraction estimates from the combined analysis of the MUSE data, corresponding to the stellar mass range indicated for the primary galaxy. The results given in this table correspond to the fractions estimated without taking into account the members of the cluster and galaxy group for the lowest redshift bin.}
\label{table:1}      % is used to refer this table in the text
\centering                          % used for centering table
\begin{tabular}{c c c c c c c }        % centered columns (11 columns)
\hline\hline  
% inserts double horizontal lines
&&&&&&\\
 $z_r$ & $\overline{z_r}$& $ \overline{M^{\star}}$ & $N_p$ & $N_g$& $f_{MM}$ \\
-&-&[log(\Msun)]&-&-&-\\
(1) & (2) & (3) & (4) & (5) & (6)  \\
&&&&&&\\
\hline                        % inserts single horizontal line
&&&&&&\\
 &&&&&&\\
 &&&$f_{\mathrm{ Major}} : 7 \leqslant $ log($M_{primary}$)$\leqslant 11$&&\\
  &&&&&&\\
   &&&&&&\\
$0.2\leqslant z< 1$&0.56& 9.63& 98 &626 &$0.078_{-0.020}^{+0.039}$\\
$1\leqslant z < 1.5$&1.26 & 9.41& 43 &352 &$0.127_{-0.047}^{+0.079}$\\
$1.5\leqslant z < 2.8$&2.07 & 9.90& 7 &141 &$0.172_{-0.051}^{+0.111}$\\
$2.8\leqslant z< 4$&3.46 & 8.57& 13 &332 &$0.081_{-0.032}^{+0.053}$\\
$4\leqslant z\leqslant 6$&5.01 & 7.72& 14 &365 &$0.075_{-0.044}^{+0.068}$ \\
&&&&&&\\
 \hline
  &&&&&&\\
 &&& $f_{\mathrm{ Major}} :7 \leqslant$log($M_{primary}$)$< 9.5$&&\\
  &&&&&&\\
 &&&&&&\\
$0.2\leqslant z< 1$&0.53&8.55& 63 &459 &$0.102_{-0.026}^{+0.038}$ \\
$1\leqslant z\leqslant 1.5$&1.25 &8.83& 30 &246 &$0.131_{-0.047}^{+0.079}$ \\
$3\leqslant z< 4$&3.47 &8.23& 11 &265 &$0.090_{-0.034}^{+0.054}$ \\
$4\leqslant z\leqslant 6$&5.03 & 7.54& 10 &272 &$0.084_{-0.046}^{+0.071}$ \\
 &&&&&&\\
 \hline
 &&&&&&\\
 &&&$f_{\mathrm{ Major}} : 9.5 \leqslant$ log($M_{primary}$)$ \leqslant 11 $&&\\
  &&&&&&\\
 &&&&&&\\
$0.2\leqslant z< 1$&0.55&10.20& 35 &152 &$0.023_{-0.016}^{+0.027}$\\
$1\leqslant z < 1.5$&1.27 & 9.73& 13 &98 &$0.172_{-0.051}^{+0.083}$\\
$1.5\leqslant z < 2.8$&2.12 & 9.90& 7 &72 &$0.255_{-0.050}^{+0.118}$\\
$2.8\leqslant z< 4$&3.49 & 9.63& 2 &70 &$0.039_{-0.024}^{+0.045}$\\
$4\leqslant z\leqslant 6$&4.93 & 10.01& 4 &81&$0.052_{-0.038}^{+0.062}$ \\
 &&&&&&\\
 \hline
   &&&&&&\\
 &&& $f_{\mathrm{ Major}} : 7 \leqslant$ log($M_{primary}$)$<$ $M_{median}(z_r) \leqslant 11$&&\\
 &&&&&&\\
  &&&&&&\\
$0.2\leqslant z< 1$&0.48& 7.92& 31 &284 &$0.088_{-0.020}^{+0.039}$\\
$1\leqslant z\leqslant 1.5$&1.20 & 8.47& 15 &150 &$0.064_{-0.041}^{+0.073}$\\
$3\leqslant z< 4$&3.45 & 7.85& 5 &140 &$0.076_{-0.031}^{+0.052}$\\
$4\leqslant z\leqslant 6$&5.10 & 7.48& 8 &141 &$0.106_{-0.052}^{+0.077}$ \\
&&&&&&\\
\hline
 &&&&&&\\
 &&& $f_{\mathrm{ Major}} : 7 \leqslant M_{median}(z_r) \leqslant$ log($M_{primary}$)$ \leqslant 11$&&\\
 &&&&&&\\
  &&&&&&\\
$0.2\leqslant z< 1$&0.60& 9.60& 67 &313 &$0.076_{-0.018}^{+0.040}$\\
$1\leqslant z < 1.5$&1.29 & 9.43& 28 &171 &$0.193_{-0.053}^{+0.085}$\\
$1.5\leqslant z < 2.8$&2.01 & 9.87& 7 &72 &$0.260_{-0.059}^{+0.119}$\\
$2.8\leqslant z< 4$&3.40 & 8.68& 8 &175 &$0.081_{-0.032}^{+0.053}$\\
$4\leqslant z\leqslant 6$&4.80 & 9.37& 6 &197 &$0.051_{-0.038}^{+0.062}$ \\
&&&&&&\\
 \hline\hline

\end{tabular}
\end{table*}

To obtain a merger fraction from a close pair count study, for each redshift bin $z_{r}$, the number of galaxy pairs, $N_p$,  must be divided by the number of primary galaxies in the parent sample, $N_g$,  and corrected from all selection effects. Indeed, observations are limited in volume and luminosity and it must be taken into account and corrected in the fraction estimates (e.g.\,de Ravel et al.\,2009).

The expression from Ventou et al.\,(2017) is used to define the major merger fraction with the addition of the new weighting scheme described in sect.\,\ref{new criteria}:
\begin{equation} 
f_{\mathrm{M}}(z_{r})=C_1\frac{\sum\limits_{K=1}^{N_p} \frac{\omega_{z}^{K_1} }{C_2(z_{r})}\frac{\omega_{z}^{K_2} }{C_2(z_{r})} W(\Delta r^P , \Delta v^P) \ \omega_{A}^K}{\sum_{i=1}^{N_g} \frac{\omega_{z}^i }{C_2(z_{r})} } 
\label{eq:merger-fraction}
\end{equation}
where $K$ is the running number attributed to a pair, $K_1$ and $K_2$ correspond to the primary and companion galaxy in the pair, respectively.
$C_1$ accounts for the missing companions due to our limit in spatial resolution at small radii (sect.\,\ref{new criteria}), and is defined as  $C_1=\frac{(r^P_{max})^{2} }{(r^P_{max})^{2}-(r^P_{min})^{2}}$. The redshift confidence weight, $\omega_{z}$, represents the confidence level associated to the spectroscopic redshift measurement:\newline
$$\omega_{z}=
\begin{cases}1 & \mbox{ if } z_{conf} =3 \text{ or } 2 , {\rm for\,secure\,redshifts}\\
0.6 & \mbox{ if } z_{conf} =1, {\rm for\,unsecure\,redshifts} \\
\end{cases}$$
The area weight takes into account the missing companion galaxy for primary galaxies on the border of the MUSE field:  $$\omega_{A} = \frac{r^P_{max}}{(r^P_{max}-r_{\rm MUSE})}$$ where $r^P_{max}$ is the radius corresponding to the projected distance limit, and $r_{\rm MUSE}$ the radius available in MUSE observations.
$W(\Delta r^P ,  \Delta v^P)$, defined in equation \ref{eq:weightbymass}, is the new weight corresponding to the probability of the close pair to merge by $z=0$ based on their relative velocity and projected separation distance.
The parameter $C_2(z_{r})$ corrects for redshift incompleteness. To compute this value we use the same method as described in Ventou et al.\,(2017) for the UDF-Mosaic and HDF-S and applied to the two other fields. Assuming that photometric redshift measurements are uniformly representative of the real redshift distribution, the number of spectroscopic redshifts is divided by the number of photometric redshifts for each bin. Photometric redshift measurements for the UDF-Mosaic field are estimated in Brinchmann et al.\,(2017) and reported in the COSMOS2015 catalog (Laigle et al., 2016) for COSMOS-Gr30 field. Photometric redshifts for A2744 were estimated using the spectral energy distribution (SED) fitting code HyperZ (Bolzonella at al.\,2000), based on photometry from the publicly available Hubble Frontier Field images of A2744 
%(ID: 13495, P.I: J. Lotz), 
in 7 filters (ACS; F435W, F606W, F814W and WFC3; F105W, F125W, F140W, F160W; Lotz et al.\,2017). A constant star formation history, a Chabrier (2003) initial mass function, a Calzetti et al.\, (2000) extinction law and templates from Bruzual \& Charlot (2003) stellar library were used as input parameters. 

Finally, uncertainties due to the cosmic variance and statistical errors on the estimated fractions are taken into account in the error budget on the merger fractions. The computed total cosmic variance for the four fields follows the recipes of Moster et al.\,(2011). A purely statistical error was derived as a confidence interval from a Bayesian approach (see e.g.\,Cameron 2011). 

Compared to Ventou et al.\,(2017), we improve our results with smaller error bars due to the increased number of galaxies in the parent and close pair samples thanks to the addition of two new observed fields A2744 and COSMOS-Gr30, as well as the new selection criteria. At first glance, it could be surprising to find similar pair fractions compared with our previous study. Indeed, in Ventou et al.\,(2017) we made the (strong) assumption that all the selected close pairs will merge by $z=0$ and thus we did not apply any weight for those pairs in the fraction estimates. With our new selection criteria, we indeed found a lower probability (between $\sim$ 30\% and 80\%) for the pairs to merge and applied a corresponding weight in the fraction estimate. However, the expected decrease of the pair fractions due to the lower probability of merging is compensated by the higher number of selected close pairs, even with a probability to merge as low as 30\%.

Figure\,\ref{fig:major_frac} shows the cosmic evolution of the major merger fraction up to $z \approx 6$, for a variety of primary galaxy stellar mass ranges. Results are summarized in Table \ref{table:1} for each redshift bin and mass range. 

We first estimate the fractions for each field individually as well as for the combined data set for major close pairs with a stellar mass primary galaxy in the range $10^7-10^{11}$\Msun\ (Fig.\,\ref{fig:major_frac}, left panel). Although the measurements are in good agreement within the error bars for the majority of the redshift bins, the impact of the environment on these estimates is clearly seen for the lowest redshift bin, $0.2\leqslant z_{r}< 1$. Due to the presence of the galaxy cluster Abell 2744 at $z\approx 0.3$ and the galaxy group in COSMOS area at $z\approx 0.7$, we observe an enhancement of the close pair counts and hence the merger fraction for these two fields compared to the UDF-Mosaic estimate. 
 Whereas we measure a major merger fraction of 21\% in A2744 and 25\% in COSMOS Gr30, which is about twice the value estimated in the UDF-Mosaic for this redshift bin, these fractions drop to 5\%  and 9\% respectively if we remove the members belonging to the galaxy cluster and to the galaxy group.

In the subsequent analysis and discussion of the merger fraction, we will restrict the samples of close pairs in the low-redshift bins by excluding those belonging to these massive structures. 
%However, this result cannot be taken at face value. 
%Although we expect to find more close pairs of galaxies in dense environments, the probability for these galaxies to effectively merge in the future should be lower in these environments due to the numerous interactions of the galaxies with other cluster/group members. 
Indeed galaxy clusters and groups provide high-density environments where near neighbors are common. However, the high velocity dispersion of low-$z$ virialized clusters and groups ($\sim 500-1000$ \kms) are not conducive to active merging among galaxies (see Mihos 2004 for a review). Indeed, measurements of the merger rate in low-redshift galaxy clusters do not generally exceed $2-3$\% (Adams et al.\,2012; Cordero et al.\,2016). Nonetheless, recent studies of high-redshift proto-clusters have shown evidence of enhanced merger rates, suggesting that merging in dense environments may play an important role in galaxy mass assembly in the early universe (Lotz et al.\,2013; Hine et al.\,2016). 
%At higher redshift $z \geq 3$, the fraction of close pairs slowly decreases with look-back time, reaching a major merger fraction of 7-8\% at $z\approx 5$.

Assuming a constant stellar mass separation of $10^{9.5}$\Msun\ for the primary galaxy, we push this analysis further and split the sample into two mass bins. Figure\,\ref{fig:major_frac} (right panel) shows the resulting evolution of the major merger fraction for massive and low-mass galaxies separately, using different merging probabilities $W(\Delta r^P ,  \Delta v^P)$ computed from equations\,\ref{eq:weightbymass}.

We observe an increase of the fraction for the high-mass sample up to 25\% at $z\approx 2$ where it reaches its maximum, followed by a decrease of the fraction down to 4-5\% between $3\leqslant z\leqslant 6$.
The major merger fraction evolution of the low-mass sample is less pronounced with a nearly flat trend, with an almost constant fraction of 8-13\% over the whole redshift range probed by our MUSE sample. Similar results are found if we consider the median mass of the parent sample in each redshift bin as the separation limit, as was done in Ventou et al.\, (2017) (see Figure \ref{fig:major_frac}, right panel). 

These evolutionary trends of the major merger fractions are in fairly good agreement with those derived from previous spectroscopic analyses (see Fig.\,\ref{fig:major_frac_2}), where they claim that beyond $z \geq 2$, the incidence of major mergers remains constant or turn over at early times (Lopez-Sanjuan et al. 2013; Tasca et al. 2014; Ventou et al. 2017). 
A comparison of the major pair fractions can also be made with recent estimates derived from the full photometric redshift probability distributions, a probabilistic approach first introduced by Lopez-Sanjuan et al. (2010). Our estimates are in very good agreement with the pair fractions derived up to $z \sim 1.5$ in the ALHAMBRA survey (Lopez-Sanjuan et al. 2015) and up to $z \sim 3.5$ in a combined analysis of UDS, VIDEO, COSMOS and GAMA surveys (Mundy et al. 2017) and recently extend to CANDELS fields (Duncan et al. 2019). All these analyses, based on photometric redshifts of large sample of galaxies, find a constant evolution of the major pair fraction with redshift as $(1+z)^n$, with a power-law index $n$ in the range $n \sim 1-3$ which is almost independent of galaxy stellar masses. The comparison at higher redshift ($z > 3$) is more tricky. Duncan et al. (2019) estimate the major pair fraction up to $z\sim 6$ for massive galaxies only, with log(\Mstar/\Msun) $> 10.3$, and find a constant rise of this fraction reaching $\sim 37$\% at $z\sim 6$. This is clearly in contradiction with our estimates ($\sim 10$\% of pairs at $z \sim 4-6$) but we need to keep in mind that the stellar mass range of galaxies probed with our MUSE deep fields is wider, extending down to much lower masses than the photometric sample used in Duncan et al. (2019).

As discussed in Ventou et al.\,(2017), predictions from cosmological simulations, such as \textsc{Horizon-AGN} (Kaviraj et al.\,2015), \textsc{Eagle} (Qu et al.\,2017) and \textsc{Illustris} (Snyder et al.\,2017), show broadly a good agreement with the cosmic evolution of the major merger fraction up to $z\sim 2-3$ (see Fig.\,\ref{fig:major_frac_2}). However, none of these simulations is making prediction above redshift $z\sim 3$. We thus compare in Fig.\,\ref{fig:major_frac_2} our results to new predictions from simulations (brown line, O'Leary et al. in prep.) which extend up to $z\sim 6$. 
These simulations employ the empirical model \textsc{Emerge} (Moster et al.\,2018) to populate dark matter halos with galaxies. The \textsc{Emerge} model utilizes a cosmological dark matter only N-body simulation in a periodic box with side lengths of $200$ Mpc. The simulations are constrained such that a suite of observations are reproduced, e.g. galaxy catalogs out to high redshift ($z\sim 6$), making it an ideal environment to study the evolving pair fraction of galaxies. Mock observations are produced from these simulated data, producing a pair fraction for each simulation snapshot. The agreement between our measurements of major pair fractions above $z\sim 3$ and these new simulations is very good. We note however that \textsc{Emerge} predicts lower values by a factor of $\sim 2$ for the major pair fractions in the redshift range $z\sim 1-3$ than those measured in the MUSE deep fields. 
%We adopt Planck $\Lambda$CDM cosmology \citep{planck} where $\Omega_m = 0.3070$, $\Omega_{\Lambda} = 0.6930$, $\Omega_b = 0.0485$, where $H_0 = 67.77\,\mathrm{km}\,\mathrm{s}^{-1}\mathrm{Mpc}^{-1}$, $n_s=0.9677$, and $\sigma_{8}=0.8149$.

\begin{figure*}[h!]
   \begin{tabular}{c}
	\includegraphics[width=1\textwidth]{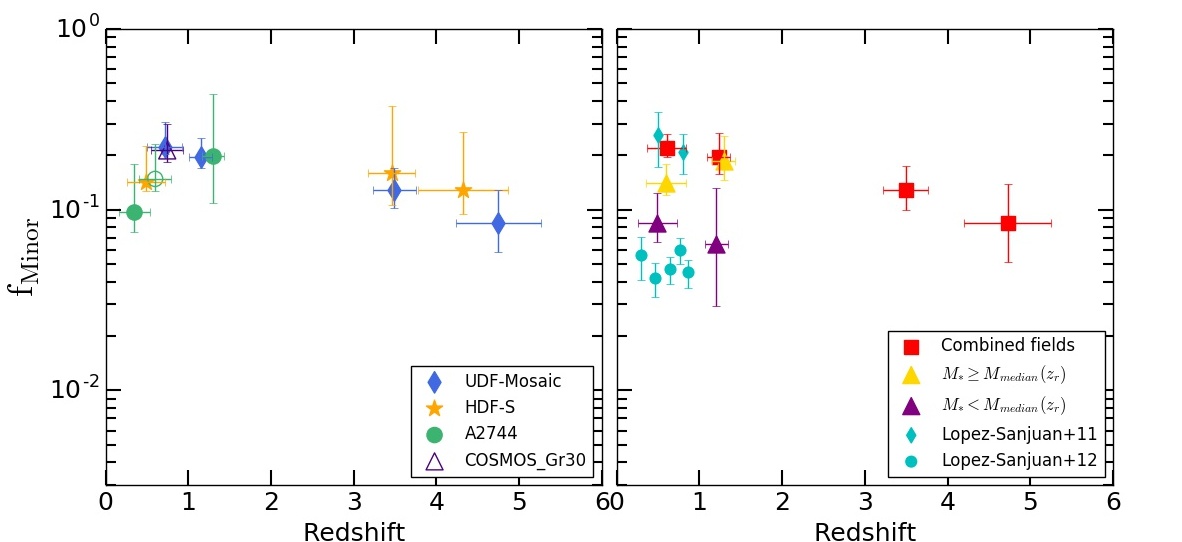}
    \end{tabular}
   \caption{Evolution of the galaxy minor merger fraction up to $z\sim 6$ from MUSE deep fields, for primary galaxies with a stellar mass range between $\sim10^{9}-10^{11}$\Msun. {\it Left}: As in Fig.\,\ref{fig:major_frac} different symbols show the results from the four regions individually. {\it Right}: red squares are estimates of the combined minor merger fraction from the whole MUSE data and over the whole stellar mass range $\sim 10^{9}-10^{11}$\Msun. Using the median value of stellar mass in each redshift bin as a separation limit ($\sim 10^{10}$\Msun, see Table\,\ref{table:2}), the purple and golden triangles correspond to the minor merger fraction for low-mass and massive galaxies respectively. Similar as in Fig.\,\ref{fig:major_frac} filled and open symbols correspond to the fractions computed without or with galaxy members of the cluster and group for $0.2 \leq z \leq 1$. Cyan points are estimates from previous spectroscopic minor pair counts (Lopez-Sanjuan et al.\,2011, 2012). }
	\label{fig:minor_frac}
\end{figure*}

\subsection{Minor merger fraction}

We derive the minor merger fraction, from the number count of galaxy close pairs with stellar mass ratio  between 1:6 and 1:100, using the same expression (equation\,\ref{eq:merger-fraction}) as for the major merger fraction. In order to keep a fairly good mass completeness in our sample, the fractions are estimated for a stellar mass range of $10^9-10^{11}$\Msun\ for the primary galaxy. 

Figure\,\ref{fig:minor_frac} shows the individual fractions for each field (left panel) and the combined data set (right panel). For A2744 and COSMOS-Gr30 only the estimates for the lowest redshift bins are shown as we are not statistically robust enough for these two fields in higher redshift intervals. For the merger fraction estimated from the combined MUSE fields, we excluded in the computations minor pairs belonging to the cluster A2744 and to the group COSMOS-Gr30. 

The minor merger fraction shows little evolution between $0.2 \leqslant z\leqslant 1.5$ with a roughly constant fraction of 20\%. Beyond $z \approx 3$, we observe a slight decrease of the fraction down to $8-13$\% in this high redshift range. Fraction estimates for the combined MUSE data set are listed in Table\,\ref{table:2}.

An attempt is made to separate the minor merger sample into two stellar mass ranges, as was done for the major close pair in section \ref{major_merger_fraction}. We take the median stellar mass in the range $10^9-10^{11}$\Msun\ of the parent sample reported in Table\,\ref{table:2} as the separation limit. For statistical reasons it is only computed for the two first redshift bins. Thus for massive primary galaxy in the range $10 \leq$ log(\Mstar/\Msun) $\leq 11$, the minor merger fraction is roughly constant around 18\% at $z\sim 0.2-1.5$ and around $6-9$\% for the low-mass sample, i.e.\,with $9 \leq$ log(\Mstar) $< 10$. 

Comparison to the few previous estimates of the minor merger fraction from spectroscopic pair counts are made in Fig.\,\ref{fig:minor_frac} (right panel).
Lopez-Sanjuan et al.\,(2012) computed the minor fraction for a mass ratio range of 1:4-1:10 and a projected separation of $ 10 \leq r^P \leq 30 $h$^{-1}$ kpc. They found a fraction around $4.5-6$\% for $z\sim 0.29-0.86$. Since their selection criteria on the minor merger sample, i.e.\, the projected distance and the mass ratio range, are narrower compared to ours, it is not surprising that our estimates of $\sim 20$\% are higher for the same redshift interval. In Lopez-Sanjuan et al.\,(2011), a minor merger fraction for bright galaxies within $r^P_{max} \sim 100 $h$^{-1}$ kpc  and a luminosity ratio in the $B-$band of 1:4$-$1:10 is reported. Their projected separation distance is more similar to this work, and their estimated fraction of 25\% and 21\% at $z=0.5$ and $z=0.8$ respectively are in good agreement with our results. 

A comparison of our minor merger fraction with recent cosmological simulations is not straightforward as the latter usually focus on the major merger fraction only and/or use different mass ratio limits to discriminate between major and minor mergers. However, we found that on average the minor merger fraction is higher than the major one by a factor of $\sim 4$, in relative good agreement with the \textsc{Horizon-AGN} simulations (Kaviraj et al\,2015) which predict a factor of $2.5-3$ between minor and major merger fractions. The difference between these two values can be explained by the different mass ratio limits to separate minor mergers from major ones. 

    \begin{table*}
\caption{Minor merger fractions up to $z\approx 6$ from MUSE deep observations for different redshift and stellar mass intervals. Same columns as in Table\ref{table:1} with Col\,(6) corresponding to the minor merger fraction estimates from the combined analysis of the MUSE data, associated to the stellar mass range indicated for the primary galaxy. These fractions are estimated without taking into account the members of the galaxy group and cluster.}
\label{table:2}      % is used to refer this table in the text
\centering                          % used for centering table
\begin{tabular}{c c c c c c c }        % centered columns (11 columns)
\hline\hline  
% inserts double horizontal lines
&&&&&&\\
 $z_r$ & $\overline{z_r}$& $ \overline{M^{\star}}$ & $N_p$ & $N_g$& $f_{mm}$ \\
-&-&[log(\Msun)]&-&-&-\\
(1) & (2) & (3) & (4) & (5) & (6)  \\
&&&&&&\\
\hline                        % inserts single horizontal line
&&&&&&\\
 &&&&&&\\
 &&& $f_{\mathrm{ Minor}} : 9 \leqslant $ log($M_{primary}$)$\leqslant 11$&&\\
 &&&&&&\\
  &&&&&&\\
$0.2\leqslant z< 1$&0.60& 10.00& 50 &260 &$0.199_{-0.032}^{+0.053}$\\
$1\leqslant z\leqslant 1.5$&1.23 & 9.85& 23 &159 &$0.196_{-0.054}^{+0.086}$\\
$2.8\leqslant z< 4$&3.49 & 9.36& 7 &100 &$0.129_{-0.041}^{+0.061}$\\
$4\leqslant z\leqslant 6$&4.73 &9.48& 10 &114 &$0.084_{-0.046}^{+0.071}$ \\
&&&&&&\\
\hline
&&&&&&\\
 &&&&&&\\
 &&& $f_{\mathrm{ Minor}} : 11 \geqslant$ log($M_{primary}$)$\geqslant M_{median}(z_r) \geqslant 9$&&\\
 &&&&&&\\
  &&&&&&\\
$0.2\leqslant z< 1$&0.59& 10.55& 22 &68 &$0.141_{-0.019}^{+0.039}$\\
$1\leqslant z\leqslant 1.5$&1.29 & 10.37& 10 &33 &$0.184_{-0.052}^{+0.084}$\\
&&&&&&\\
\hline
&&&&&&\\
 &&&&&&\\
 &&& $f_{\mathrm{ Minor}} : 9 \leqslant$ log($M_{primary}$)$<$ $M_{median}(z_r) \leqslant 11$&&\\
 &&&&&&\\
  &&&&&&\\
$0.2\leqslant z< 1$&0.49& 9.64& 28 &153 &$0.085_{-0.018}^{+0.040}$\\
$1\leqslant z\leqslant 1.5$&1.20 & 9.61& 13 &96 &$0.064_{-0.041}^{+0.073}$\\
 &&&&&&\\
 \hline\hline

\end{tabular}
\end{table*}

\section{Summary and conclusion}
\label{conclusion}

Using the \textsc{Illustris} cosmological simulation project, we investigated the relation between the velocity-distance relative separation of galaxies in a close pair and the probability that these galaxies will merge by $z=0$. We propose a new set of selection criteria for galaxy close pair counts, along with a new weighing scheme to be applied to the merger fraction. This takes into account the probability of merging for the pair derived from their relative velocity and projected separation distance.

%An investigation, using Illustris cosmological simulations, of the relation between the velocity-distance relative separation of a close pair of galaxies and the probability that these galaxies will merge by $z=0$, reveals a new set of selection criteria for galaxy close pair counts along with a new weighing scheme applied to the merger fraction which takes into account the probability of merging for the pair derived from their relative velocity and projected separation distance. 
We found that combining constraints on the projected separation distance in the sky plane and the rest-frame relative velocity of $\Delta r^P \leqslant 50$  kpc with $\Delta v^P \leqslant  300$  \kms\ and  $ 50 \leqslant \Delta r^P \leqslant  100 $ kpc with $\Delta v^P \leqslant 100$  \kms\ allows the selection of all close pairs with at least 30\% of probability to merge.

Deep MUSE observations in the HUDF, HDF-S, A2744 and COSMOS-Gr30 fields are used to construct a large spectroscopic sample of 2483 galaxies. Applying the new selection criteria, 366 secure close pairs of galaxies spread over a large redshift range ($0.2<z<6$) and stellar masses ($10^7-10^{11}$\Msun) were identified.
We use stellar masses derived from SED fitting to distinguish between major, minor and very minor close pairs using their mass ratio as proxy. We end up with a sample of 183 major, 140 minor and 47 very minor close pairs with a respective galaxy mass ratio limit of 1:6, 1:100, and lower than 1:100.

Splitting the redshift domain into five  intervals, we probe the evolution of the major and minor merger fractions up to $z\approx 6$. We leave aside the very minor close pairs which are close to the regime of smooth gaz accretion. We observe an increase of the major pair fraction in A2744 and COSMOS-Gr30 with respect to lower-density fields (HUDF and HDF-S) at $z<1$ due to the presence of the cluster ($z\sim 0.3$) and galaxy group ($z\sim 0.7$) at these redshifts. The pairs found in these two dense structures are then removed for the analysis of the merger fractions.

The sample is further divided into two ranges of stellar masses using a constant separation limit of $10^{9.5}$\Msun. Estimates for the high-mass galaxy sample show an increase of the major merger fraction up to $z\approx 2-3$ reaching a fraction of 21\% and a decrease at high redshift dropping to $\sim 5$\% at $z\approx 6$. The fraction for lower mass primary galaxies ($M_* \leq 10^{9.5}$\Msun) seems to follow a more constant evolutionary trend along cosmic time. Similar trends are found for a median stellar mass separation. 

Although we trace more accurately the merger fraction with the new criteria, the results are similar to our previous analysis over the HUDF and HDF-S fields (Ventou et al.\,2017), especially taking into account the error bars. However, error bars are narrower due to the increased number of galaxies in the parent and close pair samples. The comparison of the major merger fraction with new predictions from the \textsc{Emerge} simulations (O'Leary et al. in prep) shows a very good agreement at high redshift ($z\sim 3-6$).

The evolution of the minor merger fraction is roughly constant around 20\% for $z<1.5$ and slightly decreases for $z \geq 3$ with a fraction of $8-13$\%. The ratio between minor and major merger fractions is in good agreement with the predictions of \textsc{Horizon-AGN} simulations (Kaviraj et al.\,2015), taking into account the different mass ratio limits used to discriminate minor pairs from major ones.

\begin{acknowledgements}
We thank the referee for providing useful and constructive comments which helped to improve the quality of this paper. We warmly thank Shy Genel for giving us access to mock catalogs of galaxies generated in the \textsc{Illustris-1} simulations, which are the basis of this publication. We also acknowledge Joseph O'Leary for providing us with new results from \textsc{Emerge} simulations in advance of publication. This work has been carried out thanks to the support of the ANR FOGHAR (ANR-13-BS05-0010-02), the OCEVU Labex (ANR-11-LABX-0060), and the A*MIDEX project (ANR-11-IDEX-0001-02) funded by the ``Investissements d'avenir'' French government programme. BE acknowledges financial support from “Programme National de Cosmologie et Galaxies” (PNCG) of CNRS/INSU, France. JB acknowledges support by FCT/MCTES through national funds by this grant UID/FIS/04434/2019 and through the Investigador FCT Contract No. IF/01654/2014/CP1215/CT0003.
\end{acknowledgements}

\nocite{*}
\bibliographystyle{aa}
\bibliography{biblio_simu.bib}

\longtab{
\begin{longtable}{r r r r r r r r r r c}
\caption{\label{table:majorpairs} Sample of major (mass ratio of 1:1$-$1:6) close pairs of galaxies identified in the four MUSE fields. Labels 1 and 2 denote the primary and secondary galaxy, respectively. Cols.\,(1) and (4): Identification number in the MUSE-based catalogues. Cols.\,(2) and (5): MUSE spectroscopic redshift. Cols.\,(3) and (6): Stellar mass in logarithmic units. Cols.\,(7) and (8): Projected separation (in kpc) and velocity difference (in \kms) between the two galaxies in the pair, respectively. Cols.\,(9): sky region observed with MUSE: UDF-Mosaic, HDFS , COSMOS group CGR30, and A2744.}\\
%\label{table:majorpairs}       % is used to refer this table in the text
%\centering                          % used for centering table
%\begin{tabular}{r r r r r r r r r r c}        % centered columns (11 columns)

\hline                 % inserts double horizontal lines
%&&&&&&&&&&\\
MUSE ID$_1$ & $z_1$ & M$^{\star}_1$ & MUSE ID$_2$ & $z_2$ & M$^{\star}_2$ & r$_p$ & $\Delta_v$ & MUSE field \\    % table heading 
$-$ & $-$ & [log(\Msun)] & $-$ & $-$ & [log(\Msun)] & [kpc] & [\kms] & $-$ \\
(1) & (2) & (3) & (4) & (5) & (6) & (7) & (8) & (9) \\ 
%&&&&&&&&&&\\
%\\
\hline                        % inserts single horizontal line
%&&&&&&&&&&\\
10  & 0.275 & 8.3 & 6368  & 0.276 & 8.8 & 23.3 & 34  & \textsf{UDF-Mosaic} \\ 
24  & 2.544 & 9.8 & 35  & 2.544 & 10.0 & 14.5 & 17  & \textsf{UDF-Mosaic} \\ 
30  & 1.096 & 8.9 & 84  & 1.096 & 8.8 & 35.7 & 55  & \textsf{UDF-Mosaic} \\ 
31  & 1.851 & 9.6 & 6668  & 1.851 & 9.3 & 49.8 & 23  & \textsf{UDF-Mosaic} \\ 
32  & 1.307 & 9.2 & 65  & 1.307 & 9.0 & 38.9 & 43  & \textsf{UDF-Mosaic} \\ 
32  & 1.307 & 9.2 & 121  & 1.306 & 8.6 & 11.7 & 72  & \textsf{UDF-Mosaic} \\ 
46  & 1.413 & 9.3 & 92  & 1.414 & 8.5 & 8.2 & 20  & \textsf{UDF-Mosaic} \\ 
65  & 1.307 & 9.0 & 121  & 1.306 & 8.6 & 42.5 & 115  & \textsf{UDF-Mosaic} \\ 
96  & 0.622 & 7.7 & 108  & 0.622 & 7.8 & 20.7 & 55  & \textsf{UDF-Mosaic} \\ 
399  & 5.137 & 7.5 & 627  & 5.136 & 7.2 & 26.2 & 55  & \textsf{UDF-Mosaic} \\ 
430  & 4.513 & 7.8 & 7197  & 4.513 & 8.2 & 30.8 & 12 & \textsf{UDF-Mosaic} \\ 
891  & 0.227 & 7.8 & 6891  & 0.227 & 7.2 & 21.2 & 35  & \textsf{UDF-Mosaic} \\ 
899  & 1.097 & 10.2 & 934  & 1.096 & 9.8 & 30.5 & 94  & \textsf{UDF-Mosaic} \\ 
939  & 1.295 & 10.1 & 991  & 1.296 & 9.9 & 38.1 & 162  & \textsf{UDF-Mosaic} \\ 
950  & 0.993 & 9.0 & 1107  & 0.993 & 8.7 & 8.3 & 20 &\textsf{UDF-Mosaic} \\ 
974  & 1.087 & 9.6 & 979  & 1.086 & 9.2 & 46.8 & 59  & \textsf{UDF-Mosaic} \\ 
976  & 0.620 & 9.1 & 1020  & 0.622 & 9.4 & 48.4 & 266  & \textsf{UDF-Mosaic} \\ 
997  & 1.041 & 8.9 & 1454  & 1.041 & 8.7 & 32.6 & 24  & \textsf{UDF-Mosaic} \\ 
999  & 1.608 & 9.9 & 1268  & 1.609 & 9.7 & 7.4 & 46  & \textsf{UDF-Mosaic} \\ 
1027  & 0.219 & 7.6 & 1167  & 0.219 & 7.1 & 16.5 & 43  & \textsf{UDF-Mosaic} \\ 
1044  & 2.028 & 10.2 & 1048  & 2.028 & 10.1 & 31.8 & 81  & \textsf{UDF-Mosaic} \\ 
1064  & 1.426 & 9.7 & 6879  & 1.427 & 9.3 & 36.0 & 112  & \textsf{UDF-Mosaic} \\ 
1065  & 0.522 & 8.2 & 1444  & 0.523 & 7.6 & 28.1 & 290  & \textsf{UDF-Mosaic} \\ 
1137  & 1.096 & 9.2 & 1153  & 1.095 & 9.5 & 42.6 & 168  & \textsf{UDF-Mosaic} \\ 
1178  & 2.691 & 9.7 & 1279  & 2.691 & 9.7 & 32.5 & 65  & \textsf{UDF-Mosaic} \\ 
1188  & 1.412 & 9.6 & 1219  & 1.413 & 9.1 & 28.0 & 118  & \textsf{UDF-Mosaic} \\ 
1267  & 1.866 & 9.6 & 6947  & 1.866 & 9.8 & 32.5 & 10  & \textsf{UDF-Mosaic} \\ 
1341  & 1.413 & 9.1 & 1373  & 1.413 & 8.9 & 9.3 & 36  & \textsf{UDF-Mosaic} \\ 
1345  & 1.095 & 8.6 & 1605  & 1.095 & 8.7 & 26.9 & 37  & \textsf{UDF-Mosaic} \\ 
1353  & 1.016 & 8.3 & 7033  & 1.016 & 8.5 & 36.9 & 41  & \textsf{UDF-Mosaic} \\ 
1459  & 5.150 & 10.7 & 6477  & 5.146 & 10.3 & 40.6 & 155  & \textsf{UDF-Mosaic} \\ 
1545  & 0.992 & 8.3 & 6991  & 0.991 & 8.3 & 19.0 & 156  & \textsf{UDF-Mosaic} \\ 
1561  & 0.733 & 7.7 & 1644  & 0.732 & 7.5 & 7.0 & 67  & \textsf{UDF-Mosaic} \\ 
1611  & 0.666 & 7.8 & 1688  & 0.665 & 7.3 & 22.9 & 150  & \textsf{UDF-Mosaic} \\ 
1678  & 1.425 & 8.8 & 7101  & 1.427 & 8.7 & 32.2 & 262  & \textsf{UDF-Mosaic} \\ 
1730  & 0.681 & 7.1 & 7079  & 0.681 & 7.0 & 37.9 & 10 & \textsf{UDF-Mosaic} \\ 
2071  & 4.930 & 9.3 & 6412  & 4.928 & 9.4 & 14.0 & 97  & \textsf{UDF-Mosaic} \\ 
2679  & 3.088 & 8.3 & 4695  & 3.086 & 8.0 & 36.2 & 139  & \textsf{UDF-Mosaic} \\ 
2757  & 5.380 & 7.9 & 5398  & 5.382 & 7.2 & 33.4 & 86  & \textsf{UDF-Mosaic} \\ 
3916  & 3.973 & 8.1 & 6492  & 3.974 & 8.7 & 44.7 & 56  & \textsf{UDF-Mosaic} \\ 
4532  & 3.438 & 8.5 & 7221  & 3.435 & 8.5 & 34.1 & 215  & \textsf{UDF-Mosaic} \\ 
4542  & 4.811 & 7.2 & 5882  & 4.811 & 6.8 & 26.2 & 14 & \textsf{UDF-Mosaic} \\ 
6302  & 3.473 & 9.2 & 6925  & 3.476 & 9.6 & 32.7 & 202  & \textsf{UDF-Mosaic} \\ 
6402  & 4.372 & 8.4 & 7311  & 4.372 & 8.5 & 20.7 & 26  & \textsf{UDF-Mosaic} \\ 
6517  & 3.432 & 8.9 & 6531  & 3.432 & 8.6 & 28.6 & 22 & \textsf{UDF-Mosaic} \\ 
6531  & 3.432 & 8.6 & 7351  & 3.433 & 8.0 & 49.7 & 73  & \textsf{UDF-Mosaic} \\ 
6923  & 3.433 & 7.5 & 7283  & 3.432 & 8.0 & 21.2 & 61  & \textsf{UDF-Mosaic} \\ 
7285  & 5.486 & 7.7 & 7353  & 5.485 & 7.5 & 33.8 & 46  & \textsf{UDF-Mosaic} \\ 
27  & 1.849 & 9.8 & 41  & 1.848 & 9.9 & 74.8 & 86  & \textsf{UDF-Mosaic} \\ 
33  & 1.415 & 9.3 & 6927  & 1.416 & 8.7 & 63.4 & 95  & \textsf{UDF-Mosaic} \\ 
72  & 1.097 & 8.8 & 84  & 1.096 & 8.8 & 79.3 & 94  & \textsf{UDF-Mosaic} \\ 
391  & 3.716 & 8.4 & 6702  & 3.715 & 8.5 & 68.0 & 52  & \textsf{UDF-Mosaic} \\ 
926  & 0.667 & 9.1 & 1118  & 0.667 & 9.2 & 82.3 & 10  & \textsf{UDF-Mosaic} \\ 
944  & 0.323 & 8.2 & 7381  & 0.324 & 7.8 & 93.5 & 64  & \textsf{UDF-Mosaic} \\ 
949  & 1.316 & 9.9 & 1156  & 1.316 & 9.5 & 73.9 & 10  & \textsf{UDF-Mosaic} \\ 
950  & 0.993 & 9.0 & 1263  & 0.993 & 8.8 & 65.8 & 12  & \textsf{UDF-Mosaic} \\ 
957  & 0.678 & 8.9 & 1237  & 0.677 & 8.4 & 71.4 & 83  & \textsf{UDF-Mosaic} \\ 
963  & 0.533 & 8.7 & 1418  & 0.533 & 8.1 & 60.0 & 84  & \textsf{UDF-Mosaic} \\ 
979  & 1.086 & 9.2 & 1410  & 1.086 & 8.4 & 97.4 & 9  & \textsf{UDF-Mosaic} \\ 
987  & 0.664 & 9.0 & 1688  & 0.665 & 8.7 & 85.9 & 74  & \textsf{UDF-Mosaic} \\ 
995  & 1.036 & 9.3 & 6939  & 1.037 & 8.8 & 72.1 & 76  & \textsf{UDF-Mosaic} \\ 
1020  & 0.622 & 9.4 & 1239  & 0.622 & 8.7 & 95.7 & 44  & \textsf{UDF-Mosaic} \\ 
1030  & 1.382 & 9.9 & 1158  & 1.382 & 9.4 & 99.2 & 11 & \textsf{UDF-Mosaic} \\ 
1107  & 0.993 & 8.7 & 1263  & 0.993 & 8.8 & 73.5 & 14  & \textsf{UDF-Mosaic} \\ 
1144  & 0.620 & 7.9 & 2227  & 0.620 & 7.7 & 56.1 & 92  & \textsf{UDF-Mosaic} \\ 
1174  & 0.424 & 7.8 & 1863  & 0.424 & 7.7 & 80.0 & 31  & \textsf{UDF-Mosaic} \\ 
1205  & 0.975 & 8.8 & 7077  & 0.975 & 8.0 & 95.9 & 20  & \textsf{UDF-Mosaic} \\ 
1224  & 0.424 & 7.6 & 6955  & 0.424 & 7.5 & 91.0 & 53  & \textsf{UDF-Mosaic} \\ 
1323  & 1.098 & 9.0 & 1365  & 1.097 & 8.7 & 98.6 & 69  & \textsf{UDF-Mosaic} \\ 
1410  & 1.086 & 8.4 & 6971  & 1.087 & 8.3 & 72.8 & 38  & \textsf{UDF-Mosaic} \\ 
1496  & 1.438 & 9.0 & 6977  & 1.438 & 9.3 & 91.8 & 31  & \textsf{UDF-Mosaic} \\ 
2350  & 5.049 & 7.8 & 6435  & 5.050 & 7.8 & 97.9 & 49  & \textsf{UDF-Mosaic} \\ 
3055  & 3.675 & 8.9 & 3426  & 3.675 & 8.7 & 82.3 & 17  & \textsf{UDF-Mosaic} \\ 
4045  & 3.496 & 7.1 & 7263  & 3.496 & 6.9 & 62.1 & 15 &\textsf{UDF-Mosaic} \\ 
6294  & 5.471 & 10.3 & 7337  & 5.472 & 9.8 & 70.1 & 25 & \textsf{UDF-Mosaic} \\ 
6967  & 0.667 & 8.0 & 7035  & 0.667 & 7.7 & 57.8 & 78  & \textsf{UDF-Mosaic} \\ 
35  & 4.612 & 10.2 & 40  & 4.613 & 10.0 & 16.0 & 32  & \textsf{COSMOS} \\ 
49  & 0.724 & 9.4 & 57  & 0.725 & 9.5 & 46.7 & 121  & \textsf{COSMOS} \\ 
51  & 2.939 & 9.9 & 64  & 2.941 & 10.0 & 28.6 & 165  & \textsf{COSMOS} \\ 
55  & 0.479 & 6.9 & 81  & 0.479 & 7.5 & 41.8 & 12 & \textsf{COSMOS} \\ 
59  & 0.723 & 9.1 & 61  & 0.724 & 9.8 & 48.1 & 132  & \textsf{COSMOS} \\ 
68  & 0.726 & 10.2 & 98  & 0.725 & 10.8 & 42.4 & 158  & \textsf{COSMOS} \\ 
72  & 0.725 & 9.2 & 76  & 0.726 & 9.1 & 36.2 & 133  & \textsf{COSMOS} \\ 
92  & 0.363 & 8.4 & 101  & 0.363 & 8.6 & 30.0 & 186  & \textsf{COSMOS} \\ 
93  & 0.725 & 9.8 & 114  & 0.724 & 9.6 & 21.1 & 135  & \textsf{COSMOS} \\ 
98  & 0.725 & 10.8 & 119  & 0.725 & 10.5 & 43.8 & 66  & \textsf{COSMOS} \\ 
99  & 1.274 & 9.7 & 109  & 1.274 & 9.9 & 21.4 & 65  & \textsf{COSMOS} \\ 
103  & 0.725 & 8.2 & 118  & 0.724 & 8.9 & 46.1 & 278  & \textsf{COSMOS} \\ 
105  & 0.727 & 10.5 & 110  & 0.727 & 10.1 & 46.0 & 64  & \textsf{COSMOS} \\ 
119  & 0.725 & 10.5 & 131  & 0.723 & 10.5 & 32.2 & 283  & \textsf{COSMOS} \\ 
168  & 0.728 & 10.8 & 177  & 0.727 & 11.0 & 35.8 & 109  & \textsf{COSMOS} \\ 
170  & 0.726 & 10.3 & 186  & 0.725 & 10.5 & 46.9 & 196  & \textsf{COSMOS} \\ 
174  & 0.729 & 10.4 & 177  & 0.727 & 11.0 & 42.3 & 270  & \textsf{COSMOS} \\ 
193  & 0.727 & 9.6 & 195  & 0.727 & 9.8 & 41.7 & 15  & \textsf{COSMOS} \\ 
57  & 0.725 & 9.5 & 93  & 0.725 & 9.8 & 75.7 & 8  & \textsf{COSMOS} \\ 
71  & 0.725 & 10.9 & 98  & 0.725 & 10.8 & 82.9 & 78  & \textsf{COSMOS} \\ 
82  & 0.727 & 9.6 & 110  & 0.727 & 10.1 & 97.0 & 27  & \textsf{COSMOS} \\ 
123  & 0.723 & 9.8 & 131  & 0.723 & 10.5 & 73.6 & 53  & \textsf{COSMOS} \\ 
170  & 0.726 & 10.3 & 182  & 0.726 & 9.6 & 86.9 & 22  & \textsf{COSMOS} \\ 
10  & 1.284 & 10.8 & 27  & 1.285 & 10.6 & 47.0 & 161  & \textsf{HDFS} \\ 
29  & 0.831 & 10.4 & 58  & 0.832 & 10.2 & 25.3 & 138  & \textsf{HDFS} \\ 
32  & 0.564 & 7.9 & 135  & 0.564 & 8.0 & 39.3 & 133  & \textsf{HDFS} \\ 
50  & 2.672 & 11.0 & 55  & 2.674 & 10.8 & 6.6 & 118  & \textsf{HDFS} \\ 
104  & 1.139 & 8.9 & 567  & 1.139 & 8.6 & 5.2 & 31  & \textsf{HDFS} \\ 
183  & 3.374 & 9.8 & 261  & 3.375 & 9.8 & 4.4 & 16  & \textsf{HDFS} \\ 
433  & 3.470 & 7.3 & 478  & 3.469 & 7.2 & 20.9 & 33  & \textsf{HDFS} \\ 
441  & 4.695 & 7.8 & 499  & 4.695 & 7.3 & 5.6 & 37  & \textsf{HDFS} \\ 
10  & 1.284 & 10.8 & 15  & 1.284 & 11.2 & 95.9 & 26  & \textsf{HDFS} \\ 
23  & 0.564 & 8.5 & 135  & 0.564 & 8.0 & 77.9 & 71  & \textsf{HDFS} \\ 
9731  & 3.551 & 7.9 & M16  & 3.551 & 7.9 & 33.9 & 26  & \textsf{A2744} \\ 
2674  & 4.728 & 7.6 & 2874  & 4.728 & 6.8 & 81.5 & 10  & \textsf{A2744} \\ 
9272  & 3.476 & 7.2 & 10382  & 3.475 & 7.4 & 89.0 & 53  & \textsf{A2744} \\ 
7721  & 3.130 & 8.7 & 7858  & 3.129 & 8.4 & 62.3 & 29  & \textsf{A2744} \\ 
9356  & 1.358 & 8.2 & 10570  & 1.358 & 8.0 & 62.1 & 12  & \textsf{A2744} \\ 
10141  & 1.162 & 7.9 & 9389  & 1.161 & 8.7 & 99.8 & 69  & \textsf{A2744} \\ 
7280  & 0.946 & 8.7 & 8235  & 0.946 & 9.4 & 91.4 & 30  & \textsf{A2744} \\ 
7883  & 0.945 & 8.3 & 6364  & 0.945 & 8.0 & 78.2 & 61  & \textsf{A2744} \\ 
8853  & 0.780 & 8.0 & 6894  & 0.780 & 7.7 & 81.6 & 50  & \textsf{A2744} \\ 
11419  & 0.322 & 8.6 & 11937  & 0.322 & 9.0 & 77.3 & 68  & \textsf{A2744} \\ 
8748  & 0.320 & 8.8 & 9727  & 0.320 & 9.2 & 78.7 & 90  & \textsf{A2744} \\ 
8729  & 0.319 & 9.9 & 9646  & 0.319 & 9.8 & 50.2 & 45  & \textsf{A2744} \\ 
11418  & 0.317 & 10.2 & 11950  & 0.317 & 10.4 & 59.4 & 91  & \textsf{A2744} \\ 
11418  & 0.317 & 10.2 & 10270  & 0.317 & 9.9 & 89.0 & 45  & \textsf{A2744} \\ 
9072  & 0.316 & 10.0 & 6298  & 0.317 & 9.5 & 96.9 & 45  & \textsf{A2744} \\ 
5339  & 0.316 & 10.2 & 6298  & 0.317 & 9.5 & 69.7 & 68  & \textsf{A2744} \\ 
2768  & 0.303 & 9.3 & 3671  & 0.303 & 9.0 & 76.7 & 14  & \textsf{A2744} \\ 
9428  & 0.300 & 10.3 & 9503  & 0.300 & 10.0 & 95.3 & 92  & \textsf{A2744} \\ 
8252  & 0.297 & 9.1 & 10032  & 0.297 & 9.2 & 88.8 & 46  & \textsf{A2744} \\ 
11644  & 0.296 & 10.3 & 10478  & 0.297 & 9.8 & 65.9 & 69  & \textsf{A2744} \\ 
11655  & 0.297 & 9.3 & 10478  & 0.297 & 9.8 & 77.2 & 23  & \textsf{A2744} \\ 
8116  & 5.775 & 7.2 & 7747  & 5.771 & 7.5 & 18.5 & 185  & \textsf{A2744} \\ 
3019  & 1.368 & 8.9 & 2982  & 1.368 & 9.0 & 36.8 & 50  & \textsf{A2744} \\ 
3019  & 1.368 & 8.9 & 2907  & 1.367 & 9.3 & 17.3 & 101  & \textsf{A2744} \\ 
3019  & 1.368 & 8.9 & 2796  & 1.366 & 8.5 & 14.7 & 240  & \textsf{A2744} \\ 
2907  & 1.367 & 9.3 & 2709  & 1.366 & 9.9 & 48.9 & 164  & \textsf{A2744} \\ 
2796  & 1.366 & 8.5 & 2787  & 1.365 & 8.8 & 15.1 & 164  & \textsf{A2744} \\ 
8971  & 1.345 & 9.3 & 9010  & 1.344 & 9.1 & 5.6 & 166  & \textsf{A2744} \\ 
6638  & 1.343 & 9.3 & 6615  & 1.343 & 8.6 & 20.3 & 25  & \textsf{A2744} \\ 
11621  & 1.340 & 8.0 & 11697  & 1.340 & 8.0 & 5.3 & 51  & \textsf{A2744} \\ 
9362  & 1.163 & 8.5 & 10358  & 1.161 & 8.8 & 41.1 & 277  & \textsf{A2744} \\ 
9327  & 1.104 & 7.8 & 9049  & 1.103 & 8.0 & 8.9 & 85  & \textsf{A2744} \\ 
11952  & 1.046 & 8.7 & 10482  & 1.045 & 8.1 & 41.4 & 175  & \textsf{A2744} \\ 
7883  & 0.945 & 8.3 & 8343  & 0.944 & 8.7 & 26.3 & 215  & \textsf{A2744} \\ 
10481  & 0.671 & 8.1 & 10676  & 0.671 & 7.5 & 37.2 & 17  & \textsf{A2744} \\ 
6381  & 0.618 & 7.5 & 7516  & 0.617 & 7.4 & 45.1 & 55  & \textsf{A2744} \\ 
7542  & 0.324 & 9.2 & 8253  & 0.323 & 8.8 & 38.6 & 113  & \textsf{A2744} \\ 
8748  & 0.320 & 8.8 & 8900  & 0.320 & 8.0 & 6.2 & 90  & \textsf{A2744} \\ 
7068  & 0.320 & 10.2 & 7344  & 0.319 & 10.3 & 17.5 & 159  & \textsf{A2744} \\ 
4556  & 0.320 & 10.2 & 6034  & 0.319 & 10.7 & 35.0 & 181  & \textsf{A2744} \\ 
4556  & 0.320 & 10.2 & 4439  & 0.319 & 9.7 & 21.8 & 227  & \textsf{A2744} \\ 
5436  & 0.318 & 8.5 & 4433  & 0.318 & 8.4 & 37.4 & 45  & \textsf{A2744} \\ 
5436  & 0.318 & 8.5 & 4580  & 0.319 & 8.0 & 43.2 & 90  & \textsf{A2744} \\ 
4439  & 0.319 & 9.7 & 3910  & 0.318 & 9.9 & 39.8 & 159  & \textsf{A2744} \\ 
4828  & 0.319 & 8.4 & 4580  & 0.319 & 8.0 & 21.9 & 45  & \textsf{A2744} \\ 
4433  & 0.318 & 8.4 & 4580  & 0.319 & 8.0 & 39.3 & 136  & \textsf{A2744} \\ 
6849  & 0.317 & 9.5 & 6298  & 0.317 & 9.5 & 38.6 & 68  & \textsf{A2744} \\ 
6849  & 0.317 & 9.5 & 6872  & 0.316 & 9.0 & 17.0 & 136  & \textsf{A2744} \\ 
6298  & 0.317 & 9.5 & 6872  & 0.316 & 9.0 & 38.3 & 68  & \textsf{A2744} \\ 
7231  & 0.292 & 8.4 & 6982  & 0.292 & 8.0 & 20.3 & 208  & \textsf{A2744} \\ 
6843  & 0.306 & 8.8 & 7609  & 0.306 & 9.4 & 17.4 & 22  & \textsf{A2744} \\ 
12443  & 0.304 & 9.9 & 12269  & 0.303 & 9.8 & 13.2 & 68  & \textsf{A2744} \\ 
8143  & 0.303 & 9.4 & 7199  & 0.303 & 8.8 & 41.3 & 207  & \textsf{A2744} \\ 
9382  & 0.303 & 10.6 & 8930  & 0.302 & 10.0 & 15.1 & 184  & \textsf{A2744} \\ 
12269  & 0.303 & 9.8 & 12149  & 0.303 & 9.8 & 46.7 & 92  & \textsf{A2744} \\ 
5134  & 0.301 & 8.8 & 5978  & 0.301 & 8.6 & 34.5 & 69  & \textsf{A2744} \\ 
8907  & 0.301 & 10.1 & 9428  & 0.300 & 10.3 & 41.5 & 161  & \textsf{A2744} \\ 
6776  & 0.301 & 8.4 & 6211  & 0.301 & 8.2 & 37.1 & 115  & \textsf{A2744} \\ 
6776  & 0.301 & 8.4 & 7214  & 0.301 & 7.8 & 16.1 & 46  & \textsf{A2744} \\ 
5576  & 0.300 & 9.3 & 6163  & 0.299 & 9.2 & 35.2 & 92  & \textsf{A2744} \\ 
5576  & 0.300 & 9.3 & 5693  & 0.299 & 9.8 & 17.0 & 299  & \textsf{A2744} \\ 
5576  & 0.300 & 9.3 & 6339  & 0.299 & 9.4 & 36.5 & 230  & \textsf{A2744} \\ 
6163  & 0.299 & 9.2 & 5693  & 0.299 & 9.8 & 49.8 & 207  & \textsf{A2744} \\ 
6163  & 0.299 & 9.2 & 6339  & 0.299 & 9.4 & 5.4 & 138  & \textsf{A2744} \\ 
5693  & 0.299 & 9.8 & 6339  & 0.299 & 9.4 & 49.9 & 69  & \textsf{A2744} \\ 
4893  & 0.297 & 6.4 & 5174  & 0.298 & 6.6 & 4.2 & 115  & \textsf{A2744} \\ 
10032  & 0.297 & 9.2 & 10703  & 0.297 & 9.1 & 49.5 & 92  & \textsf{A2744} \\ 
10703  & 0.297 & 9.1 & 12079  & 0.296 & 8.4 & 40.9 & 161  & \textsf{A2744} \\ 
9876  & 0.294 & 9.5 & 10243  & 0.293 & 9.5 & 23.5 & 208  & \textsf{A2744} \\ 
7291  & 0.293 & 7.2 & 7042  & 0.292 & 6.5 & 6.5 & 162  & \textsf{A2744} \\

% &&&&&&&&&&\\

\hline                                   %inserts single line
%\end{tabular}
\end{longtable}
}

\longtab{
\begin{longtable}{r r r r r r r r r r c}
\caption{\label{table:minorpairs} Sample of minor (mass ratio of 1:6$-$1:100) close pairs of galaxies identified in the four MUSE fields. See Table.\,\ref{table:1} for column references.}\\
%\label{table:minorpairs}      % is used to refer this table in the text
%\centering                          % used for centering table
%\begin{tabular}{r c c r r c c r r r c}        % centered columns (11 columns)
\hline                % inserts double horizontal lines
%&&&&&&&&&&\\
MUSE ID$_1$ & $z_1$ & M$^{\star}_1$ & MUSE ID$_2$ & $z_2$ & M$^{\star}_2$ & r$_p$ & $\Delta_v$ & MUSE field \\    % table heading 
$-$ & $-$ & [log(\Msun)] & $-$ & $-$ & [log(\Msun)] & [kpc] & [\kms] & $-$ \\
(1) & (2) & (3) & (4) & (5) & (6) & (7) & (8) & (9) \\ 
%&&&&&&&&&&\\
%\\
\hline                        % inserts single horizontal line
%&&&&&&&&&&\\
1  & 0.622 & 10.4 & 1004  & 0.622 & 9.2 & 49.7 & 79  & \textsf{UDF-Mosaic} \\ 
4  & 0.765 & 10.0 & 14  & 0.765 & 9.1 & 9.5 & 19  & \textsf{UDF-Mosaic} \\ 
8  & 1.095 & 10.3 & 72  & 1.097 & 8.8 & 26.3 & 243  & \textsf{UDF-Mosaic} \\ 
265  & 3.886 & 9.3 & 633  & 3.885 & 7.4 & 6.8 & 66  & \textsf{UDF-Mosaic} \\ 
357  & 3.436 & 8.4 & 6666  & 3.439 & 9.8 & 46.2 & 198  & \textsf{UDF-Mosaic} \\ 
412  & 4.136 & 7.8 & 6698  & 4.136 & 8.9 & 19.2 & 12  & \textsf{UDF-Mosaic} \\ 
430  & 4.513 & 7.8 & 6342  & 4.517 & 8.9 & 4.0 & 224  & \textsf{UDF-Mosaic} \\ 
861  & 0.151 & 8.9 & 954  & 0.151 & 7.6 & 12.4 & 38  & \textsf{UDF-Mosaic} \\ 
869  & 0.665 & 10.0 & 1243  & 0.665 & 8.2 & 36.6 & 98  & \textsf{UDF-Mosaic} \\ 
874  & 0.458 & 9.8 & 906  & 0.458 & 8.8 & 9.3 & 12  & \textsf{UDF-Mosaic} \\ 
889  & 0.620 & 9.9 & 1131  & 0.621 & 8.2 & 36.4 & 289  & \textsf{UDF-Mosaic} \\ 
899  & 1.097 & 10.2 & 1108  & 1.095 & 8.7 & 37.8 & 284  & \textsf{UDF-Mosaic} \\ 
924  & 1.098 & 10.3 & 6965  & 1.100 & 8.8 & 33.5 & 217  & \textsf{UDF-Mosaic} \\ 
934  & 1.096 & 9.8 & 1108  & 1.095 & 8.7 & 26.5 & 190  & \textsf{UDF-Mosaic} \\ 
943  & 0.663 & 9.2 & 1346  & 0.663 & 8.1 & 45.2 & 26  & \textsf{UDF-Mosaic} \\ 
948  & 0.735 & 9.8 & 1465  & 0.735 & 7.8 & 43.0 & 43  & \textsf{UDF-Mosaic} \\ 
950  & 0.993 & 9.0 & 1993  & 0.994 & 7.6 & 38.8 & 140  & \textsf{UDF-Mosaic} \\ 
959  & 0.907 & 9.4 & 6949  & 0.907 & 8.5 & 28.3 & 17  & \textsf{UDF-Mosaic} \\ 
965  & 0.113 & 7.1 & 1862  & 0.113 & 5.7 & 27.5 & 24  & \textsf{UDF-Mosaic} \\ 
974  & 1.087 & 9.6 & 6971  & 1.087 & 8.3 & 33.6 & 13  & \textsf{UDF-Mosaic} \\ 
979  & 1.086 & 9.2 & 6971  & 1.087 & 8.3 & 29.9 & 45  & \textsf{UDF-Mosaic} \\ 
981  & 1.095 & 9.2 & 1280  & 1.095 & 8.2 & 8.5 & 35  & \textsf{UDF-Mosaic} \\ 
982  & 1.096 & 9.7 & 6450  & 1.095 & 8.7 & 42.9 & 137  & \textsf{UDF-Mosaic} \\ 
1011  & 1.035 & 10.5 & 1062  & 1.035 & 9.7 & 9.6 & 10  & \textsf{UDF-Mosaic} \\ 
1017  & 1.438 & 10.2 & 6977  & 1.438 & 9.3 & 35.0 & 19  & \textsf{UDF-Mosaic} \\ 
1107  & 0.993 & 8.7 & 1993  & 0.994 & 7.6 & 35.8 & 140  & \textsf{UDF-Mosaic} \\ 
1156  & 1.316 & 9.5 & 1575  & 1.315 & 8.3 & 43.8 & 75  & \textsf{UDF-Mosaic} \\ 
1459  & 5.150 & 10.7 & 6507  & 5.145 & 8.8 & 19.2 & 203  & \textsf{UDF-Mosaic} \\ 
1504  & 3.607 & 9.4 & 6878  & 3.609 & 7.7 & 16.1 & 134  & \textsf{UDF-Mosaic} \\ 
1538  & 1.094 & 8.5 & 7382  & 1.094 & 7.7 & 20.4 & 53  & \textsf{UDF-Mosaic} \\ 
1615  & 1.288 & 7.9 & 1655  & 1.288 & 8.9 & 10.9 & 43  & \textsf{UDF-Mosaic} \\ 
1639  & 1.099 & 7.9 & 1908  & 1.099 & 8.9 & 32.9 & 44  & \textsf{UDF-Mosaic} \\ 
1835  & 4.810 & 10.0 & 6249  & 4.811 & 11.5 & 36.7 & 46  & \textsf{UDF-Mosaic} \\ 
2532  & 0.752 & 7.5 & 6519  & 0.753 & 8.7 & 22.6 & 97  & \textsf{UDF-Mosaic} \\ 
3621  & 3.068 & 8.3 & 6451  & 3.069 & 7.4 & 10.5 & 72  & \textsf{UDF-Mosaic} \\ 
3952  & 3.416 & 7.7 & 6921  & 3.416 & 8.6 & 11.0 & 23  & \textsf{UDF-Mosaic} \\ 
6295  & 5.135 & 6.3 & 6327  & 5.135 & 7.5 & 32.8 & 15  & \textsf{UDF-Mosaic} \\ 
6342  & 4.517 & 9.2 & 7197  & 4.513 & 8.2 & 34.7 & 222  & \textsf{UDF-Mosaic} \\ 
6477  & 5.146 & 10.3 & 6507  & 5.145 & 8.8 & 21.7 & 48  & \textsf{UDF-Mosaic} \\ 
6517  & 3.432 & 8.9 & 7351  & 3.433 & 8.0 & 43.6 & 76  & \textsf{UDF-Mosaic} \\ 
7033  & 1.016 & 8.5 & 7083  & 1.017 & 7.6 & 16.6 & 143  & \textsf{UDF-Mosaic} \\ 
8  & 1.095 & 10.3 & 30  & 1.096 & 8.9 & 54.5 & 93  & \textsf{UDF-Mosaic} \\ 
16  & 1.097 & 10.0 & 89  & 1.096 & 8.7 & 99.6 & 25  & \textsf{UDF-Mosaic} \\ 
17  & 0.844 & 9.3 & 7145  & 0.844 & 7.8 & 96.0 & 95  & \textsf{UDF-Mosaic} \\ 
96  & 0.622 & 7.7 & 1004  & 0.622 & 9.2 & 98.4 & 86  & \textsf{UDF-Mosaic} \\ 
879  & 1.087 & 10.8 & 974  & 1.087 & 9.6 & 88.9 & 99  & \textsf{UDF-Mosaic} \\ 
884  & 0.737 & 9.8 & 6516  & 0.737 & 8.8 & 55.0 & 48  & \textsf{UDF-Mosaic} \\ 
919  & 1.096 & 9.5 & 1345  & 1.095 & 8.6 & 80.2 & 61  & \textsf{UDF-Mosaic} \\ 
919  & 1.096 & 9.5 & 1605  & 1.095 & 8.7 & 97.2 & 24  & \textsf{UDF-Mosaic} \\ 
925  & 1.294 & 10.2 & 1404  & 1.294 & 8.8 & 67.4 & 28  & \textsf{UDF-Mosaic} \\ 
936  & 0.666 & 9.2 & 1695  & 0.665 & 7.8 & 87.3 & 42  & \textsf{UDF-Mosaic} \\ 
974  & 1.087 & 9.6 & 1410  & 1.086 & 8.4 & 54.2 & 51  & \textsf{UDF-Mosaic} \\ 
1017  & 1.438 & 10.2 & 1496  & 1.438 & 9.0 & 68.2 & 21  & \textsf{UDF-Mosaic} \\ 
1056  & 3.072 & 11.9 & 7051  & 3.071 & 10.0 & 65.0 & 83  & \textsf{UDF-Mosaic} \\ 
1121  & 0.734 & 8.5 & 2478  & 0.734 & 7.0 & 54.3 & 21  & \textsf{UDF-Mosaic} \\ 
1835  & 4.810 & 10.0 & 7343  & 4.809 & 8.0 & 98.1 & 47  & \textsf{UDF-Mosaic} \\ 
2074  & 4.833 & 9.5 & 2865  & 4.833 & 7.7 & 98.5 & 11  & \textsf{UDF-Mosaic} \\ 
2116  & 3.468 & 9.5 & 2515  & 3.469 & 8.6 & 63.2 & 58  & \textsf{UDF-Mosaic} \\ 
2679  & 3.088 & 8.3 & 6030  & 3.088 & 6.6 & 86.4 & 10 & \textsf{UDF-Mosaic} \\ 
2755  & 4.497 & 7.9 & 7269  & 4.498 & 8.7 & 81.6 & 59  & \textsf{UDF-Mosaic} \\ 
6392  & 4.472 & 9.1 & 7287  & 4.471 & 7.5 & 75.5 & 45  & \textsf{UDF-Mosaic} \\ 
6915  & 3.704 & 7.2 & 7239  & 3.704 & 8.9 & 87.5 & 67  & \textsf{UDF-Mosaic} \\ 
7047  & 4.229 & 8.2 & 7157  & 4.230 & 9.1 & 71.8 & 50  & \textsf{UDF-Mosaic} \\ 
2077  & 4.780 & 7.6 & 2264  & 4.779 & 6.6 & 57.4 & 93  & \textsf{A2744} \\ 
10725  & 3.476 & 8.5 & 10669  & 3.476 & 9.5 & 78.9 & 13  & \textsf{A2744} \\ 
10472  & 1.162 & 9.7 & 9389  & 1.161 & 8.7 & 95.5 & 41  & \textsf{A2744} \\ 
8184  & 0.944 & 9.5 & 6371  & 0.944 & 7.8 & 76.1 & 77  & \textsf{A2744} \\ 
8253  & 0.323 & 8.8 & 7954  & 0.323 & 10.2 & 63.6 & 45  & \textsf{A2744} \\ 
8748  & 0.320 & 8.8 & 10059  & 0.320 & 10.6 & 79.9 & 9 & \textsf{A2744} \\ 
7230  & 0.320 & 9.5 & 10059  & 0.320 & 10.6 & 86.3 & 45  & \textsf{A2744} \\ 
8900  & 0.320 & 8.0 & 9727  & 0.320 & 9.2 & 78.7 & 9  & \textsf{A2744} \\ 
5436  & 0.318 & 8.5 & 4439  & 0.319 & 9.7 & 54.5 & 90  & \textsf{A2744} \\ 
5436  & 0.318 & 8.5 & 3910  & 0.318 & 9.9 & 89.4 & 68  & \textsf{A2744} \\ 
4433  & 0.318 & 8.4 & 3910  & 0.318 & 9.9 & 55.0 & 22  & \textsf{A2744} \\ 
9072  & 0.316 & 10.0 & 6872  & 0.316 & 9.0 & 61.5 & 22  & \textsf{A2744} \\ 
6043  & 0.316 & 9.5 & 6527  & 0.316 & 10.4 & 84.5 & 68  & \textsf{A2744} \\ 
6210  & 0.315 & 10.5 & 6892  & 0.315 & 9.7 & 90.4 & 22  & \textsf{A2744} \\ 
4423  & 0.303 & 10.5 & 3671  & 0.303 & 9.0 & 58.1 & 69  & \textsf{A2744} \\ 
8907  & 0.301 & 10.1 & 6776  & 0.301 & 8.4 & 83.1 & 69  & \textsf{A2744} \\ 
6776  & 0.301 & 8.4 & 9428  & 0.300 & 10.3 & 89.8 & 92  & \textsf{A2744} \\ 
7824  & 0.300 & 10.8 & 5576  & 0.300 & 9.3 & 69.7 & 23  & \textsf{A2744} \\ 
6339  & 0.299 & 9.4 & 8117  & 0.299 & 10.3 & 98.8 & 10  & \textsf{A2744} \\ 
10314  & 0.297 & 10.1 & 8252  & 0.297 & 9.1 & 93.0 & 13  & \textsf{A2744} \\ 
12404  & 5.054 & 8.7 & 12026  & 5.054 & 7.8 & 3.1 & 9  & \textsf{A2744} \\ 
4926  & 4.336 & 9.4 & 5574  & 4.334 & 7.5 & 25.0 & 106  & \textsf{A2744} \\ 
7721  & 3.130 & 8.7 & 7701  & 3.131 & 7.2 & 3.6 & 101  & \textsf{A2744} \\ 
2504  & 1.766 & 9.5 & 2532  & 1.766 & 8.4 & 4.3 & 18  & \textsf{A2744} \\ 
3019  & 1.368 & 8.9 & 2709  & 1.366 & 9.9 & 31.9 & 265  & \textsf{A2744} \\ 
2907  & 1.367 & 9.3 & 2796  & 1.366 & 8.5 & 9.3 & 139  & \textsf{A2744} \\ 
2709  & 1.366 & 9.9 & 2787  & 1.365 & 8.8 & 31.0 & 139  & \textsf{A2744} \\ 
6090  & 1.344 & 10.5 & 6638  & 1.343 & 9.3 & 32.3 & 51  & \textsf{A2744} \\ 
6090  & 1.344 & 10.5 & 6615  & 1.343 & 8.6 & 34.9 & 76  & \textsf{A2744} \\ 
10472  & 1.162 & 9.7 & 10358  & 1.161 & 8.8 & 33.0 & 55  & \textsf{A2744} \\ 
8235  & 0.946 & 9.4 & 7883  & 0.945 & 8.3 & 27.3 & 123  & \textsf{A2744} \\ 
7542  & 0.324 & 9.2 & 7954  & 0.323 & 10.2 & 35.3 & 158  & \textsf{A2744} \\ 
8748  & 0.320 & 8.8 & 8729  & 0.319 & 9.9 & 47.9 & 272  & \textsf{A2744} \\ 
8748  & 0.320 & 8.8 & 9646  & 0.319 & 9.8 & 14.1 & 227  & \textsf{A2744} \\ 
10059  & 0.320 & 10.6 & 9727  & 0.320 & 9.2 & 44.5 & 90  & \textsf{A2744} \\ 
8900  & 0.320 & 8.0 & 9646  & 0.319 & 9.8 & 10.1 & 136  & \textsf{A2744} \\ 
4556  & 0.320 & 10.2 & 4828  & 0.319 & 8.4 & 31.7 & 181  & \textsf{A2744} \\ 
6034  & 0.319 & 10.7 & 4439  & 0.319 & 9.7 & 37.7 & 45  & \textsf{A2744} \\ 
4439  & 0.319 & 9.7 & 4828  & 0.319 & 8.4 & 41.5 & 45  & \textsf{A2744} \\ 
4439  & 0.319 & 9.7 & 4433  & 0.318 & 8.4 & 31.2 & 136  & \textsf{A2744} \\ 
4439  & 0.319 & 9.7 & 4580  & 0.319 & 8.0 & 23.3 & 17 & \textsf{A2744} \\ 
4538  & 0.315 & 9.0 & 5061  & 0.314 & 10.3 & 23.9 & 205  & \textsf{A2744} \\ 
11418  & 0.317 & 10.2 & 11531  & 0.317 & 8.7 & 39.4 & 136  & \textsf{A2744} \\ 
11531  & 0.317 & 8.7 & 11950  & 0.317 & 10.4 & 30.7 & 45  & \textsf{A2744} \\ 
9072  & 0.316 & 10.0 & 7367  & 0.315 & 8.9 & 45.6 & 296  & \textsf{A2744} \\ 
7231  & 0.292 & 8.4 & 7291  & 0.293 & 7.2 & 11.6 & 162  & \textsf{A2744} \\ 
7231  & 0.292 & 8.4 & 7042  & 0.292 & 6.5 & 5.3 & 15  & \textsf{A2744} \\ 
7229  & 0.304 & 10.0 & 6814  & 0.304 & 8.8 & 15.5 & 114  & \textsf{A2744} \\ 
10239  & 0.301 & 8.9 & 10689  & 0.301 & 10.5 & 30.8 & 161  & \textsf{A2744} \\ 
7824  & 0.300 & 10.8 & 6163  & 0.299 & 9.2 & 44.0 & 69  & \textsf{A2744} \\ 
7824  & 0.300 & 10.8 & 6339  & 0.299 & 9.4 & 39.1 & 207  & \textsf{A2744} \\ 
5136  & 0.299 & 5.7 & 5174  & 0.298 & 6.6 & 14.2 & 184  & \textsf{A2744} \\ 
10314  & 0.297 & 10.1 & 10032  & 0.297 & 9.2 & 48.7 & 46  & \textsf{A2744} \\ 
11644  & 0.296 & 10.3 & 11655  & 0.297 & 9.3 & 49.5 & 92  & \textsf{A2744} \\ 
7291  & 0.293 & 7.2 & 7251  & 0.293 & 5.5 & 4.7 & 115  & \textsf{A2744} \\ 
7251  & 0.293 & 5.5 & 7042  & 0.292 & 6.5 & 9.5 & 46  & \textsf{A2744} \\ 
6982  & 0.292 & 8.0 & 7042  & 0.292 & 6.5 & 25.4 & 208  & \textsf{A2744} \\ 
57  & 0.725 & 9.5 & 69  & 0.726 & 8.4 & 31.0 & 79  & \textsf{COSMOS} \\ 
61  & 0.724 & 9.8 & 71  & 0.725 & 10.9 & 25.4 & 119  & \textsf{COSMOS} \\ 
82  & 0.727 & 9.6 & 98  & 0.725 & 10.8 & 40.7 & 290  & \textsf{COSMOS} \\ 
57  & 0.725 & 9.5 & 98  & 0.725 & 10.8 & 74.3 & 35  & \textsf{COSMOS} \\ 
59  & 0.723 & 9.1 & 131  & 0.723 & 10.5 & 53.8 & 19  & \textsf{COSMOS} \\ 
68  & 0.726 & 10.2 & 69  & 0.726 & 8.4 & 91.4 & 72  & \textsf{COSMOS} \\ 
69  & 0.726 & 8.4 & 93  & 0.725 & 9.8 & 51.4 & 88  & \textsf{COSMOS} \\ 
142  & 0.724 & 8.9 & 176  & 0.725 & 10.6 & 51.0 & 43  & \textsf{COSMOS} \\ 
6  & 0.422 & 9.5 & 101  & 0.422 & 7.5 & 31.6 & 11  & \textsf{HDFS} \\ 
40  & 3.012 & 8.8 & 56  & 3.008 & 10.9 & 15.4 & 292  & \textsf{HDFS} \\ 
50  & 2.672 & 11.0 & 51  & 2.673 & 9.4 & 22.8 & 61  & \textsf{HDFS} \\ 
51  & 2.673 & 9.4 & 55  & 2.674 & 10.8 & 17.6 & 57  & \textsf{HDFS} \\ 
65  & 2.020 & 10.4 & 125  & 2.019 & 8.8 & 45.4 & 123  & \textsf{HDFS} \\ 
159  & 3.745 & 10.1 & 386  & 3.742 & 9.0 & 28.5 & 197  & \textsf{HDFS} \\ 
238  & 3.820 & 10.5 & 514  & 3.823 & 8.5 & 45.6 & 199  & \textsf{HDFS} \\ 
3  & 0.564 & 9.8 & 23  & 0.564 & 8.5 & 99.3 & 68  & \textsf{HDFS} \\ 
28  & 0.318 & 8.0 & 566  & 0.318 & 6.5 & 82.6 & 20  & \textsf{HDFS} \\ 
35  & 1.281 & 10.6 & 64  & 1.281 & 9.1 & 72.1 & 96  & \textsf{HDFS} \\ 
202  & 3.277 & 8.2 & 449  & 3.277 & 9.8 & 57.9 & 53  & \textsf{HDFS} \\ 
216  & 4.018 & 9.7 & 308  & 4.018 & 8.0 & 69.7 & 52  & \textsf{HDFS} \\
% &&&&&&&&&&\\
\hline                                   %inserts single line
%\end{tabular}
\end{longtable}
}

\appendix
\section{A new weighting scheme for the merger fraction}
\label{reg}

\begin{figure}
	\includegraphics[width=\columnwidth]{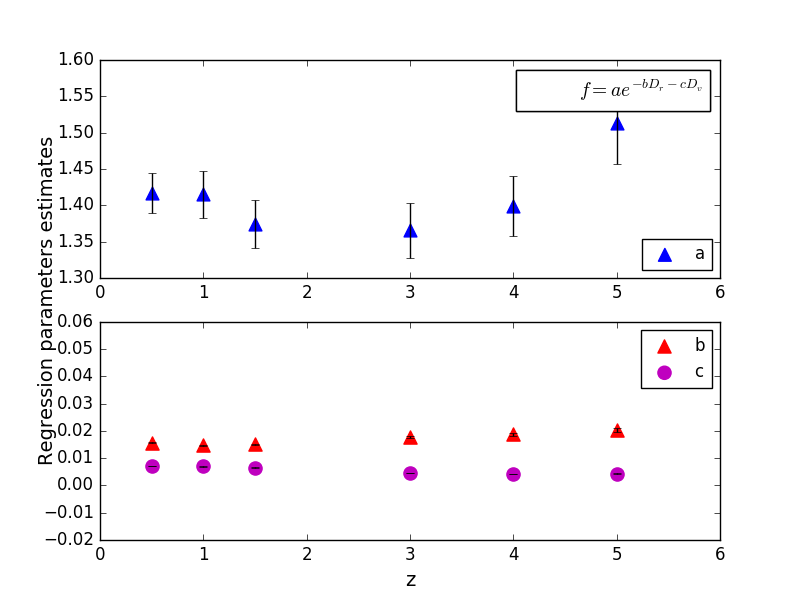}
   \caption{Redshift evolution of the regression parameters estimates with the reported approximate function. The error bars represent the computed 1$\sigma$ errors on the parameters.}
	\label{reg2}
\end{figure}

\begin{figure*}[!h]
    \begin{supertabular}{c}
       \includegraphics[width=1\textwidth]{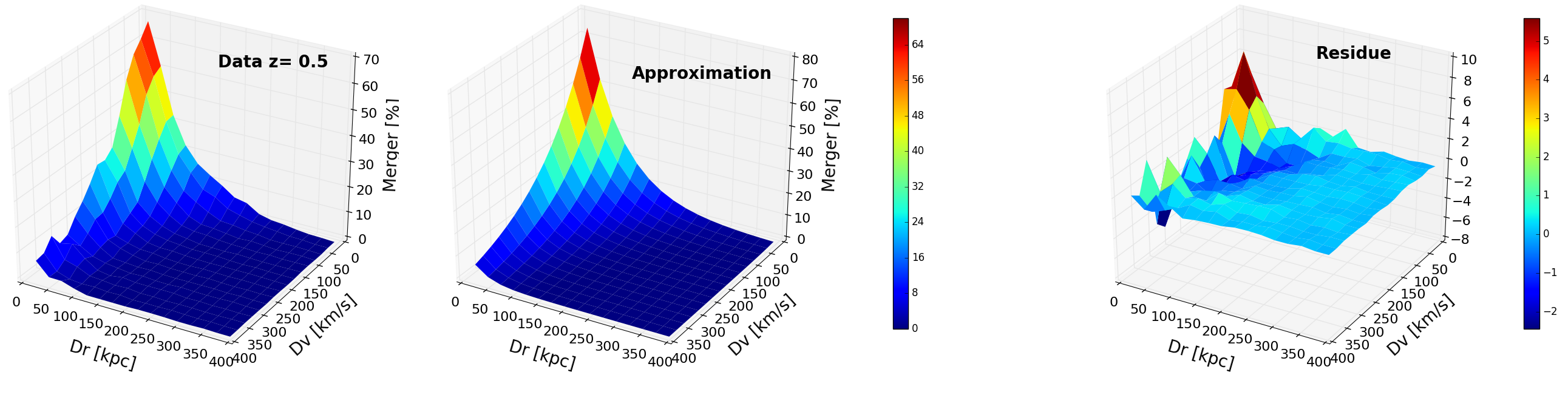}\\
 
        \includegraphics[width=1\textwidth]{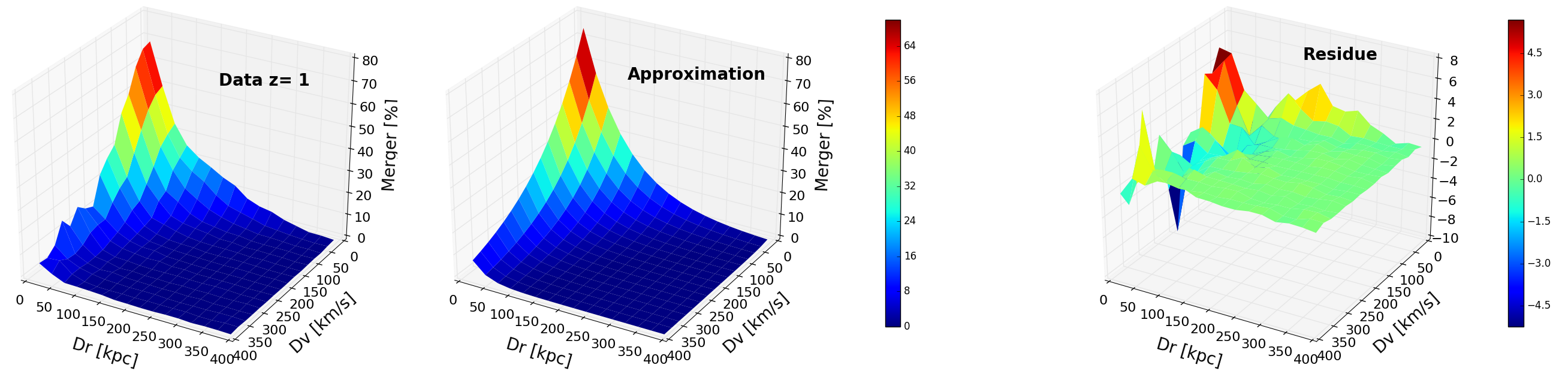} \\
 
        \includegraphics[width=1\textwidth]{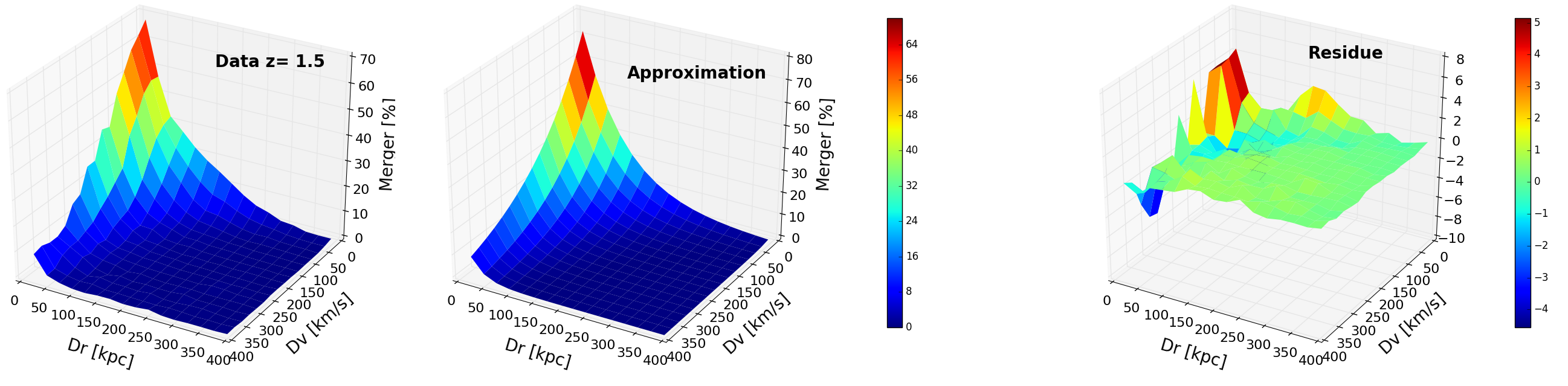} \\
        
        \includegraphics[width=1\textwidth]{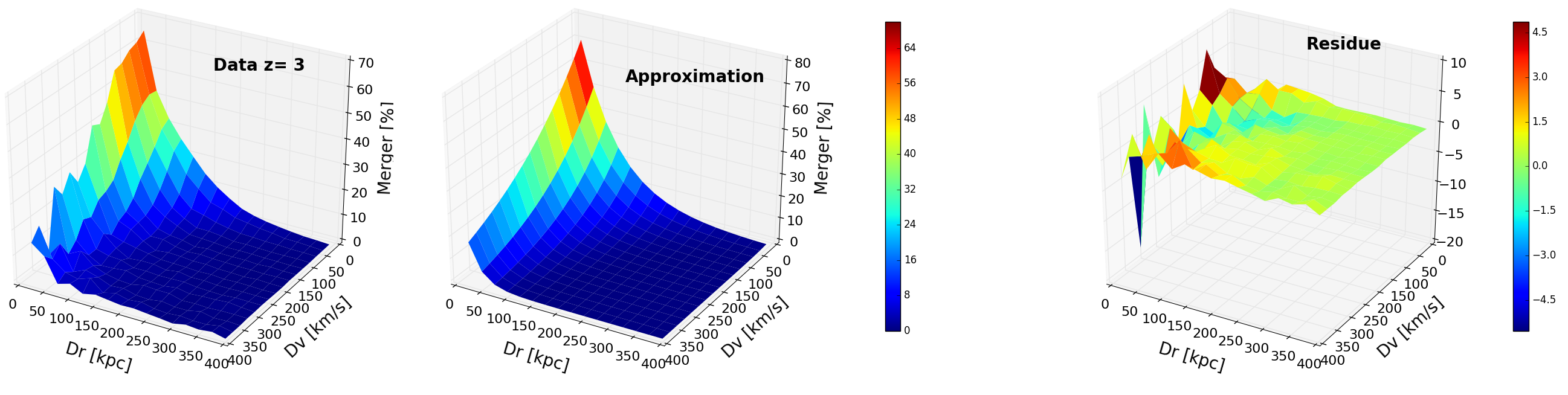} \\  
        \includegraphics[width=1\textwidth]{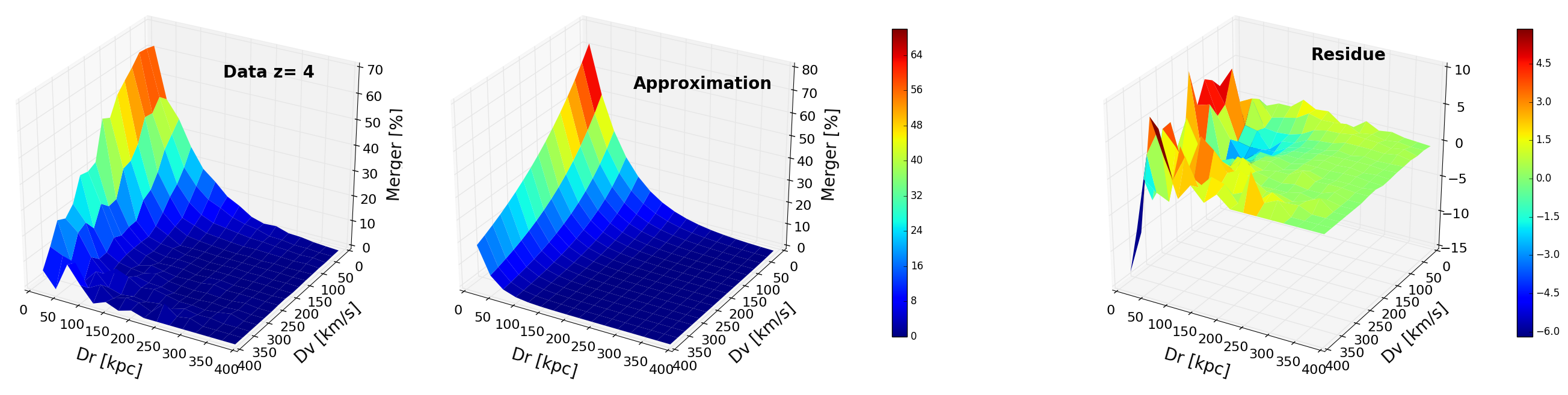} \\
 \end{supertabular}\end{figure*}
  \newpage\newpage
\begin{figure*}[!h]
    \begin{supertabular}{c}    
        \includegraphics[width=1\textwidth]{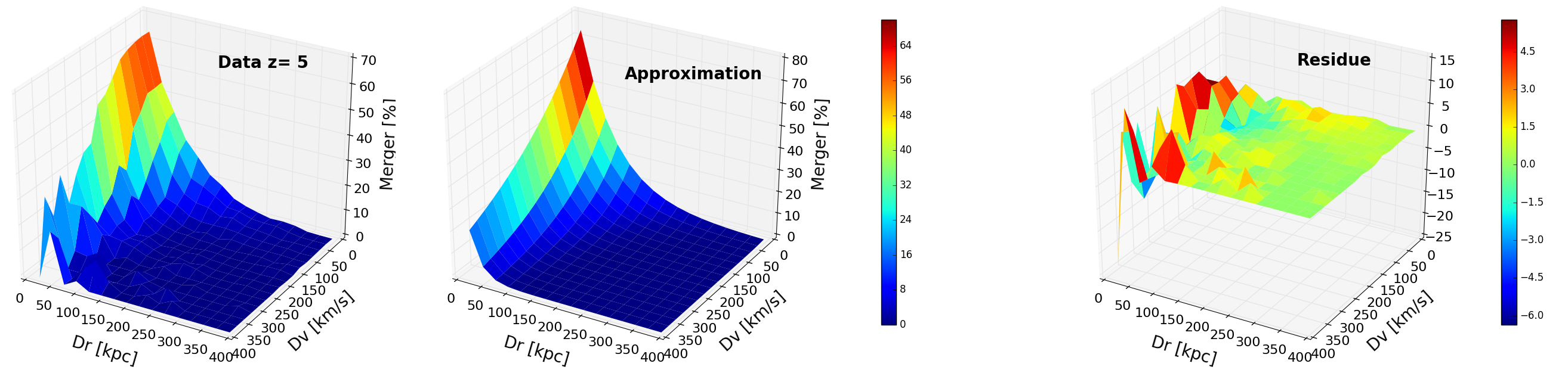} \\
        \end{supertabular}
\caption{{\it From left to right} : For each redshift, the projected velocity-separation distance diagram, the approximate function from the nonlinear regression and the residual of their subtraction.}
\label{reg1}
\end{figure*}

In Fig~\ref{reg1} we compare, for the six redshift snapshots, the projected velocity-separation distance diagrams obtained in section \ref{criteria} to the least-squares fits of an exponential function using a non-linear regression to the simulated datasets. The redshift evolution of the parameter estimates is shown in Fig~\ref{reg2}. We decide to use the median of the parameter estimates in the final expression of the probability weight, $W(\Delta r^P,  \Delta v^P)$, effective for all redshift (see equation~\ref{eq:weight}), since there is little evolution of the different parameters with redshift.

\section{Catalogs of close pairs of galaxies detected in MUSE fields}
\label{properties}

\end{document}